\documentclass[aps,ams,prc,superscriptaddress,showpacs,floatfix]{revtex4-1}
\usepackage{epsfig}
\usepackage{graphicx}
\usepackage{amsmath}
\usepackage{multirow}
\usepackage{graphicx}
\usepackage{amsmath}
\usepackage{feynmf}
\usepackage[normalem]{ulem}  
\usepackage[dvips]{color} 

\renewcommand\sout{\bgroup \color{red} \ULdepth=-.5ex \ULset}

\newcommand{\Ex}[2]{\ifmmode{#1\times10^{#2}}\else{$#1\times10^{#2}$}\fi}

\begin{document}
\title{QCD sum rules for the neutron, $\Sigma$, and $\Lambda$ in neutron matter}

\author{Kie Sang Jeong}
\email[]{k.s.jeong@yonsei.ac.kr}
\email[]{kiesang.jeong@apctp.org}
\author{Giju Gye}
\email[]{kkj8907@yonsei.ac.kr}
\author{Su Houng Lee}
\email[]{suhoung@yonsei.ac.kr}
\affiliation{Department of Physics
and Institute of Physics and Applied Physics, Yonsei University,
Seoul 120-749, Republic of Korea}

\date{Revised 2 October 2016)}
\begin{abstract}
The nuclear density dependencies of the neutron and $\Sigma$ and $\Lambda$
hyperons are important inputs in the determination of the neutron star
mass as the appearance of hyperons coming from  strong attractions significantly changes the
stiffness of the equation of state (EOS) at iso-spin asymmetric
dense nuclear matter.
 In-medium spectral sum
rules have been analyzed for the nucleon, $\Sigma$, and $\Lambda$
hyperon to investigate their properties up to slightly above the saturation
nuclear matter density by using the linear density approximation for the condensates. The construction scheme of the
interpolating fields without derivatives has been reviewed and used
to construct a general interpolating field for each baryon with
parameters specifying the strength of independent interpolating
fields. Optimal choices for the interpolating fields were obtained
by requiring the sum rules to be stable against variations of the
parameters and  the result to be consistent with known
phenomenology.
 The optimized result shows that Ioffe's choice is not suitable for the $\Lambda$ hyperon sum rules.  It is found that, for the $\Lambda$ hyperon interpolating
field, the up and down quark combined into the scalar diquark
structure $u^T C \gamma_5 d$ should be emphasized to ensure stable
sum rules. The quasi-$\Sigma$ and -$\Lambda$ hyperon energies are
always found to be higher than the quasineutron energy in the
region $0.5 <\rho/\rho_0<1.5 $ where the linear density
approximation in the sum-rule analysis is expected to be reliable.
\end{abstract}

\maketitle
\section{Introduction}
Observations of 2$M_\odot$ neutron star~\cite{Demorest, Antoniadis}
sparked a renewed interest for the EOS of dense nuclear matter at
large iso-spin asymmetry~\cite{Fraga:2013qra, Drago:2014oja,
Drews:2014spa}. In the low-density limit, the nuclear matter can be
regarded as a gas of weakly interacting quasiparticles filled up to
their respective quasi-Fermi sea.  In such a limit, the early
appearance of the hyperon will make the matter soft because the
additional degrees of freedom can be filled in the matter without
enhancing the quasi-Fermi sea.  Model calculations tend to show that
such a soft EOS will not support a 2$M_\odot$ neutron
star~\cite{Glendenning:1984jr, Knorren:1995ds}.
 Therefore, it is generally believed that the
hyperon degrees of freedom will eventually become repulsive at high dense matter due to
the interactions between nucleon and hyperon, which can lead to a
stiff EOS at high  nuclear matter density.
 Even if the
nuclear matter is strongly correlated, the appearance of hyperon degrees
of freedom usually reduces the energy density of the matter and the
maximum mass of the neutron star becomes bounded to values smaller than
2$M_\odot$~\cite{Glendenning:1984jr, Knorren:1995ds}. Hence, the
density behavior of the nucleon as well as the hyperon are of great
current interest.

As for the nucleon, the density dependence the quasineutron energy
in the asymmetric nuclear matter is characterized by the nuclear
symmetry energy~\cite{Li:2008gp}.
The nuclear matter energy density will have an additional variation axis if the  matter includes a nontrivial fraction of hyperons.
 Unfortunately, it is hard to
calculate the EOS of the dense matter in the hadron phase directly
from first principle or from models with relevant effective degrees of freedom because experimental information on these is scare at present. Hopefully, worldwide plans for rare
isotope machines that can probe the symmetry energy may improve the
situation. Until more experimental data are available, effective
method based on quantum chromodynamics (QCD) can be an alternative
approach to directly calculate the properties of quasibaryons  at
high density.

The operator-product-expansion (OPE-based) spectral sum rule (QCD sum
rules) is a well established method for investigating the properties
of hadrons~\cite{Shifman:1978bx, Ioffe:1981kw, Reinders:1984sr}.
Through the OPE in QCD degrees of freedom, the nonperturbative QCD
contribution at confined phase can be systematically included into
the spectral structure of hadron resonance in both the vacuum and
nuclear medium. Using theoretical estimates and the
results of experimental measurements for the nuclear expectation values of condensates, QCD
sum-rules methods have been successfully applied to study nucleons~\cite{Drukarev:1988kd, Cohen:1991js, Furnstahl:1992pi,
Jin:1992id, Jin:1993up, Cohen:1994wm} and vector
mesons~\cite{Hatsuda:1991ez, Klingl:1997kf, Leupold:1997dg,
Klingl:1998sr} in the nuclear medium. However, a systematic QCD sum-rule
study for the in-medium properties of the $\Sigma$ and $\Lambda$ hyperons
using a generalized interpolating fields and updated condensate
values is still missing. Although the sum rules have been updated by
using some of the condensate values extracted from theoretical development including lattice
QCD~\cite{Alarcon:2012nr, Ren:2014vea, Durr:2015dna, Yang:2015uis}, additional
assumptions  have to be made for the result to be consistent with
existing experimental observations from the $\Lambda$ and $\Sigma$
hyper-nuclei~\cite{Millener:1988hp, Noumi:2001tx}.

In previous studies, the interpolating fields for hyperons have
been obtained via SU(3) flavor transformation from the well
established Ioffe choice for the nucleon interpolating
fields~\cite{Shifman:1978bx, Ioffe:1981kw, Reinders:1984sr,
Jin:1993fr, Jin:1994bh}. But in principle, an independent basis for
interpolating fields can be freely constructed as long as it has the
required quantum numbers of the hadron of interest. Therefore, the
generalized interpolating fields would be a linear combination of
the basis set with parameters specifying the strength of the independent
basis. An optimal interpolating field will reflect the dominant
quark configuration of the ground state and thus have a large
overlap with the ground state. If the parameter set are close to the
optimal choice, the sum rule would be stable under small variations
in the parameters, whereas an unfavorable choice will be reflected
in an unstable sum rule.

In this study, we review the construction scheme of the
interpolating fields without derivatives for the neutron and $\Sigma$
and $\Lambda$ hyperons and constructed  a general interpolating field
for each baryon with parameters specifying the strength of
independent bases. Optimal choices for the interpolating fields were
obtained by requiring the sum rules to be stable against variations
of the parameters and the result to be consistent with known
phenomenology. The self-energies and the energy of the quasibaryon
states and their density behavior have been calculated by the QCD sum-rules approach with the renewed interpolating fields.

This paper is organized as follows: in Sec.~\ref{sec2}, a brief
introduction for in-medium QCD sum rules and arguments for
constructing interpolating fields are presented. In Sec.~\ref{sec3},
detailed OPE for correlation functions are given, and the treatment
for in medium condensates are presented. In Sec.~\ref{sec4}, the sum
rules for the neutron, $\Sigma$, and $\Lambda$ baryons in the neutron
matter are analyzed. Discussion and conclusions are given in
Sec.~\ref{sec5}.

\section{QCD sum rules and interpolating fields}\label{sec2}

In the Bjorken limit, the scattering amplitude of a hadron and
leptons can be calculated by the OPE of the correlation function between
explicit quark currents, which overlaps with the partonic
configuration in the hadron. This means that the quantum number of a
hadron can be interpolated with explicit quark current from the QCD
Lagrangian.  Moreover, the OPE calculation in the QCD degrees of
freedom is justified in this limit. One can also study the properties of
a hadron by constructing the corresponding spectral sum rules by
using the OPE of the current-current correlation function which
contains the hadron state as the ground state. This type of approach
is based on the short distance expansion and requires following
assumptions:

\begin{itemize}
 \item[1.] \emph{The interpolating fields should be constructed  to have the quantum
number of the hadron of interest and chosen to have a strong overlap
with it.}
 \item[2.] \emph{Among the states that appear in the correlator, the
hadron state should be a well separated ground state.}
\end{itemize}

The correlation function of the baryon interpolating fields is
defined as
\begin{align}
\Pi(q) \equiv i \int d^4 x e^{iqx} \langle\Psi_0\vert
\textrm{T}[\eta(x)\bar{\eta}(0)]\vert\Psi_0\rangle,\label{corr}
\end{align}
where $\eta(x)$ is an interpolating field for the baryon and
$\vert\Psi_0\rangle$ is the parity and time-reversal symmetric
ground state, either of the vacuum or the medium. In the medium
case, the state is characterized by its density
$\rho=\rho_n+\rho_p$, the matter velocity $u_\mu$, and the iso-spin
asymmetry factor $I=(\rho_n-\rho_p)/(\rho_n+\rho_p)$. With assumed
parity, time-reversal symmetry, and Lorentz covariance, the
correlation function can be decomposed into three
invariants~\cite{Furnstahl:1992pi}:
\begin{align}
\Pi(q) \equiv \Pi_s(q^2,qu)+\Pi_q(q^2,qu) \slash \hspace{-0.2cm}
q+\Pi_u(q^2,qu) \slash \hspace{-0.2cm}u. \label{corrd}
\end{align}
The medium is taken to be at rest; $u^\mu\rightarrow(1,\vec{0})$
and $\Pi_i(q^2,q
u)\rightarrow\Pi_i(q_0,\vert\vec{q}\vert\rightarrow\textrm{fixed})$.
Each invariant satisfies the following dispersion relation on the complex
$\omega$ plane:
\begin{align}
\Pi_i(q_0,\vert\vec{q}\vert)&=\frac{1}{2\pi i}\int^\infty_{-\infty}
d\omega  \frac{ \Delta\Pi_i(\omega,\vert\vec{q}\vert)}{\omega-q_0}
+F_n(q_0 ,\vert\vec{q}\vert), \label{dis-eo}
\end{align}
where $F _n(q_0 ,\vert\vec{q}\vert) \equiv
F^e_n(q_0^2,\vert\vec{q}\vert) +q_0F^o_n(q_0^2,\vert\vec{q}\vert)$
is a finite-order polynomial. The discontinuity
$\Delta\Pi_i(\omega,\vert\vec{q}\vert)$ can be defined as follows:
\begin{align}
\Delta\Pi_i(\omega,\vert\vec{q}\vert)
&\equiv\lim_{\epsilon\rightarrow0^{+}}
[\Pi_i(\omega+i\epsilon,\vert\vec{q}\vert)-\Pi_i
(\omega-i\epsilon,\vert\vec{q}\vert)]=2i\textrm{Im}[\Pi_i(\omega+i\epsilon,\vert\vec{q}\vert)]
\nonumber\\
&=\Delta\Pi^e_i(\omega^2,\vert\vec{q}\vert)+\omega
\Delta\Pi^o_i(\omega^2,\vert\vec{q}\vert).\label{discm}
\end{align}
All the possible resonances including the ground state (quasibaryon
pole) are contained in the discontinuity \eqref{discm}. By using
these relations, the invariants can be decomposed into an even and an odd
part of $q_0$, where each part has the following dispersion relation at
fixed $\vert\vec{q}\vert$:
\begin{align}
\Pi_i(q_0,\vert\vec{q}\vert)&=\Pi_i^e(q_0^2,\vert\vec{q}\vert)+q_0
\Pi_i^o(q_0^2,\vert\vec{q}\vert),\\
\Pi_i^e(q_0^2,\vert\vec{q}\vert)& =\frac{1}{2\pi
i}\int^\infty_{-\infty} d\omega \frac{\omega^2
}{\omega^2-q_0^2}\Delta\Pi^o_i(\omega^2,\vert\vec{q}\vert)
+F^e_n(q_0^2,\vert\vec{q}\vert),\label{dise}
  \\
\Pi_i^o(q_0^2,\vert\vec{q}\vert)& =\frac{1}{2\pi
i}\int^\infty_{-\infty} d\omega \frac{
 1}{\omega^2-q_0^2}\Delta\Pi^e_i(\omega^2,\vert\vec{q}\vert)
+F^o_n(q_0^2,\vert\vec{q}\vert),\label{diso}
\end{align}
where $\Delta\Pi_i^e(q_0^2,\vert\vec{q}\vert)$ and
$\Delta\Pi_i^o(q_0^2,\vert\vec{q}\vert)$ are even functions of
$q_0$.

If one regards the baryon in the nuclear medium as a quasiparticle
state, the nuclear interaction can be accounted for as the in-medium
self-energies within the mean-field potential. The phenomenological
structure of the correlation function near the quasibaryon pole can
then be suggested as
\begin{align}
\Pi(q)&\Rightarrow \frac{\lambda^{*2}}{\slash
\hspace{-0.2cm}q-M_B-\Sigma(q)}=\lambda^{*2}\frac{\slash
\hspace{-0.2cm}q+ M_B^{*}- \Sigma_v \slash
\hspace{-0.2cm}u}{(q_0-E_q)(q_0-\bar{E}_q)},\label{medpole}
\end{align}
where $ \Sigma(q)=\Sigma_s(q^2,qu)+\Sigma_v(q)\slash \hspace{-0.2cm}
u$, $ M_B^{*} = M_B+\Sigma_s(q^2,qu)$,
$E_q=\Sigma_v+(\vec{q}^2+M_B^{*2})^{\frac{1}{2}}$,
$\bar{E}_q=\Sigma_v-(\vec{q}^2+M_B^{*2} )^{\frac{1}{2}}$ and $\lambda^{*}$ is
the residue at the quasibaryon pole, which accounts for the overlap
between the interpolating fields and the quasi-baryon state.
 The invariants can then be
identified as follows:
\begin{align}
\Pi_s(q_0,\vert\vec{q}\vert)&=-\lambda^{*2}\frac{M_B^{ *}}{(q_0-E_q)(q_0-\bar{E}_q)}+\cdots,\\
\Pi_q(q_0,\vert\vec{q}\vert)&=-\lambda^{*2}\frac{1}{(q_0-E_q)(q_0-\bar{E}_q)}+\cdots,\\
\Pi_u(q_0,\vert\vec{q}\vert)&=\lambda^{*2}\frac{\Sigma_v}{(q_0-E_q)(q_0-\bar{E}_q)}+\cdots.
\end{align}

On the other hand, the invariants can be expressed in terms of QCD
degrees of freedom by OPE in following limit: $q^2 \rightarrow -
\infty$, $\vert\vec{q}\vert\rightarrow\textrm{fixed}$ (equivalent to
$q_0^2 \rightarrow - \infty$,
$\vert\vec{q}\vert\rightarrow\textrm{fixed}$):
\begin{align}
\Pi_i(q^2,q_0^2) &=\sum_n C^i_n(q^2,q_0^2)\langle
\hat{O}_n\rangle_{\rho,I},
\end{align}
where $C^i_n(q^2,q_0^2)$ are the Wilson coefficients, and the condensate
part $\langle \hat{O}_n\rangle_{\rho,I}$ has been evaluated within the
linear density approximation.

Depending on the quantum number and the intrinsic structure  of the
ground-state hadron, construction of the interpolating fields should
be different. In the following section, we argue that while Ioffe's
choice can be suitable for describing the nucleon and $\Sigma$
hyperon family, it is not so for the $\Lambda$ hyperon.

\subsection{Interpolating fields for the $\Sigma$ hyperon}

\begin{figure}
\includegraphics[height=2.7cm]{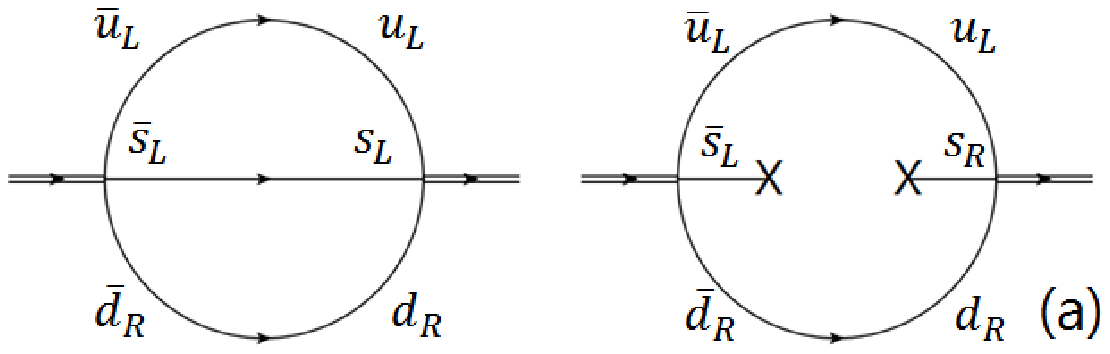}
\includegraphics[height=2.7cm]{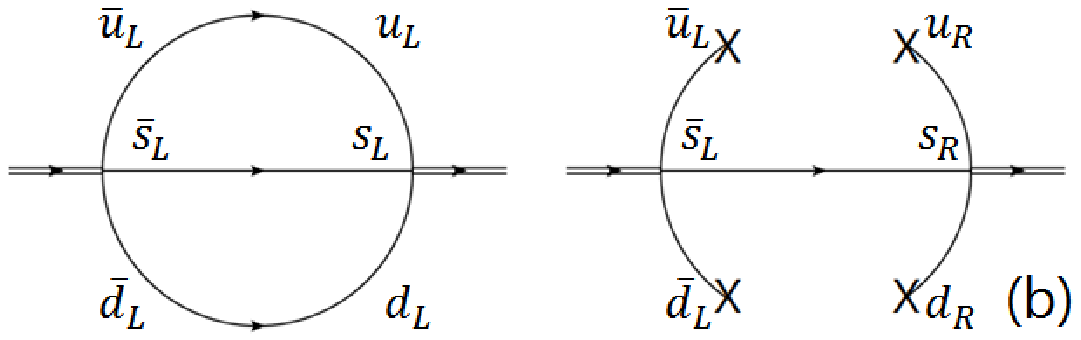}
\caption{Diagrammatic description for the $\Sigma^{0}$ correlation
function. Set (a) shows the self correlation function of basis
\eqref{scrnt1}. Set (b) shows the self correlation function of basis
\eqref{scrnt2}. In each set, the left diagram shows the partonic
propagation over short distance and the right diagram shows the
possible lowest dimensional quark condensate contribution. Only
strange quarks can propagate to the different helicity state from the
initial helicity state.}\label{currentc}
\end{figure}

For the baryon interpolating fields, the most simple structure
can be composed by a diquark without derivatives and  an attached
external quark that carries the fermionic nature of the baryon. If
one requires the diquarks to be composed of light quarks without any
derivatives, they can be classified into either the iso-spin-asymmetric ($I=0$) or -symmetric ($I=1$) configuration. The set of
interpolating fields with the diquark in $I=0$ configuration can be
written as follows:
\begin{align}
\left\{q_1,q_2~\textrm{in}~I=0\right\}=\left\{\epsilon_{abc}[q_{1a}^T
C q_{2b}]\gamma_5 q_{3c},~\epsilon_{abc}[q_{1a}^T C \gamma_5 q_{2b}]
q_{3c},~\epsilon_{abc}[q_{1a}^T C \gamma_5 \gamma_\mu
q_{2b}]\gamma^\mu q_{3c}\right\},\label{di-asym}
\end{align}
where $q_1$ and $q_2$ stand for the light quark flavors and $q_3$
stands for the external quark flavor. The set for $I=1$
configuration can be written as follows:
\begin{align}
\left\{q_1,q_2~\textrm{in}~I=1\right\}=\left\{\epsilon_{abc}[q_{1a}^T
C \gamma_\mu q_{2b}]\gamma_5 \gamma^\mu
q_{3c},~\epsilon_{abc}[q_{1a}^T C \sigma_{\mu\nu} q_{2b}]\gamma_5
\sigma^{\mu\nu}q_{3c}\right\}.\label{di-sym}
\end{align}
Hence, for the $\Sigma$ family, the interpolating fields can be
expressed as linear combinations of the bases in the
set~\eqref{di-sym} with $q_3$ taken to be the external strange quark
flavor.

On the other hands, the interpolating fields can also be constructed
by requiring (i) the diquark structure in the $s=0$ configuration without any
derivative and (ii) the light quarks in the $I=1$ configuration. Either
way, the most general lowest dimensional interpolating fields for
the $\Sigma$ can be written as follows:
\begin{align}
\eta_{\Sigma(t)}&=\epsilon_{abc}\left([q^T_{1a}C s_b]\gamma_5
q_{2c}+[q^T_{2a} C s_b]\gamma_5
q_{1c}+t\left([q^T_{1a} C \gamma_5 s_b] q_{2c}+[q^T_{2a} C \gamma_5 s_b] q_{1c}\right)\right)\nonumber\\
&=\left(\frac{1-t}{2}\right)\epsilon_{abc}[q^T_{1a} C\gamma_\mu
q_{2b}]\gamma_5\gamma^\mu
s_c+\left(\frac{1+t}{4}\right)\epsilon_{abc}[q^T_{1a}
C\sigma_{\mu\nu} q_{2b}]\gamma_5 \sigma^{\mu\nu} s_c,\label{scrnt}
\end{align}
where Fierz rearrangement has been executed to make the diquark
parts carry the iso-spin information ($I=1$) and the strange quark
carry the spin information ($s=1/2$). In the second line, the
independent basis set reduced to the set~\eqref{di-sym}. The
corresponding fields for $\Sigma^{+}$, $\Sigma^{0}$, and $\Sigma^{-}$
can be obtained by choosing $q_1$ and $q_2$ for the appropriate
light quark flavor. After choosing $q_1=u$ and $q_2=d$ for
$\Sigma^{0}$, each basis can be expressed in the helicity states:
\begin{align}
\epsilon_{abc}[u^T_aC\gamma_\mu d_b]\gamma_5\gamma^\mu s_c&=
2\epsilon_{abc} \left( [u^T_{R,a}C s_{R,b}]d_{L,c}+[d^T_{R,a}C s_{R,b}]u_{L,c} - (L \leftrightarrow R) \right)  \label{scrnt1},\\
\epsilon_{abc}[u^T_aC\sigma_{\mu\nu} d_b]\gamma_5\sigma^{\mu\nu}
s_c&= 4\epsilon_{abc} \left( [u^T_{R,a}C s_{R,b}]d_{R,c}+[d^T_{R,a}C
s_{R,b}]u_{R,c} - (L \leftrightarrow R) \right)\label{scrnt2},
\end{align}
where the subscripts $L$ and $R$ denote the left- and right-helicity
states, respectively.

Both bases can contribute to the short-ranged partonic propagation
(perturbative contribution). On the other hand, if one tries to
include the long-ranged correlation (nonperturbative contribution),
then the second basis \eqref{scrnt2} may have a problem at the
lowest order. To understand this issue consider the propagation
between two helicity states. Near the separation scale of the OPE,
for which the mass scale of the low-lying baryon ($\sim
1~\textrm{GeV}$) is taken, the light quark mass is negligible.
Hence, for the light quarks, the perturbative propagation preserves
helicity. However, because the strange quark mass is non-negligible
compared with the separation scale, the strange quark propagation can
mix helicity: the helicity mixing  part is proportional to the
strange quark mass. In the nonperturbative regime, the leading chiral
symmetry-breaking term occurs as $\langle \bar{q}_L q_R\rangle $ or
$\langle \bar{q}_R q_L\rangle $, hence occurs only between
correlations of mixed helicity. This mechanism explains the origin of the vacuum baryon masses~\cite{Ioffe:1981kw} and has the following
effect for the correlation function:

In the OPE of the self correlation function of the basis
\eqref{scrnt1}, the lowest dimensional quark condensate is $\langle
\bar{s} s \rangle $ [Fig.~\ref{currentc}(a)]. The medium part of
$\langle \bar{s} s \rangle $ can be estimated in the linear density
approximation with recent lattice QCD studies \cite{Durr:2015dna,
Yang:2015uis}. However, as one can see in Fig.~\ref{currentc}(b), in
the OPE of the self-correlation function of the basis
\eqref{scrnt2},  the four-quark condensates, whose matrix element is
still not  known well, appear as the lowest-dimensional quark
condensate. The cross correlation function between basis
\eqref{scrnt1} and \eqref{scrnt2} also cannot have two-quark
condensate as the lowest mass dimensional term. If one knows the
value of dimension-6 four-quark condensates, inclusion of basis
\eqref{scrnt2} will not be any problem, because the physical results
should not depend on the choice of the basis. However, because we
have only limited constraints to the four-quark condensates,  the
uncertainty of these expectation values will be amplified if they
appear as the leading quark operator. Setting $t=-1$ (Ioffe's
choice), one can suppress the contribution from the self-correlation
function of basis \eqref{scrnt2} and consequently avoid the problem.

In Ioffe's interpolating fields \eqref{scrnt1}, the diquark
structure is a pseudovector $u^T C\gamma_\mu d $ ($s^P=0^{-}$ for
$\mu=0$, $s^P=1^{+}$ for $\mu=i$). In spatial rotation, the time
component behaves as a scalar ($s=0$) and spatial components behave as
a three-vector ($s=1$). Therefore, the relative angular momentum
between the light quarks in the nonrelativistic quasi-$\Sigma$
state described through the basis \eqref{scrnt1} should be in the $l=1$
state.

\subsection{Interpolating fields for the nucleon}

A similar argument can be made for the nucleon case. In this
case, the most simple structure would be composed by (i) the diquark
in $I=0$ configuration and (ii) an attached external quark which
carries the fermionic nature and iso-spin of the nucleon. The linear
combination can be written as
\begin{align}
\eta_{N(t_1,t_2)}&=\epsilon_{abc}\left([q_{1a}^T C q_{2b}]\gamma_5
q_{3c}+t_1[q_{1a}^T C \gamma_5 q_{2b}] q_{3c} +t_2[q_{1a}^T C
\gamma_5 \gamma_\mu q_{2b}]\gamma^\mu q_{3c}\right),\label{nbas}
\end{align}
where the light quark flavors $q_1$ and $q_2$ are in $I=0$
configuration and $q_3=u$ ($q_3=d$) for the proton (neutron). In the
nucleon case, $q_1=q_3 \neq q_2$ or $q_2=q_3 \neq q_1$ and the light
quarks with  the same flavor should be in the $I=1$ configuration: the
third basis in the interpolating field \eqref{nbas} can be
rearranged as
\begin{align}
\epsilon_{abc}[q_{1a}^T C \gamma_5 \gamma_\mu q_{2b}]\gamma^\mu
q_{1c}=-\frac{1}{2}\epsilon_{abc}[q_{1a}^T C \gamma_\mu
q_{1b}]\gamma_5 \gamma^\mu q_{2c}=-\epsilon_{abc}\left([q_{1a}^T C
q_{2b}]\gamma_5 q_{1c}-[q_{1a}^T C \gamma_5 q_{2b}] q_{1c}\right),
\end{align}
so that  the number of independent bases in the interpolating fields
\eqref{nbas} has been reduced to two:
\begin{align}
\eta_{N(t)}&=2\epsilon_{abc}\left([q_{1a}^T C q_{2b}]\gamma_5
q_{1c}+t[q_{1a}^T C \gamma_5 q_{2b}] q_{1c} \right)\nonumber\\
&=\left(\frac{1-t}{2}\right)\epsilon_{abc}[q_{1a}^T C\gamma_\mu
q_{1b}]\gamma_5\gamma^\mu
q_{2c}+\left(\frac{1+t}{4}\right)\epsilon_{abc}[q_{1a}^T
C\sigma_{\mu\nu} q_{1b}]\gamma_5 \sigma^{\mu\nu}
q_{2c}.\label{ncrnt}
\end{align}
After choosing $q_1=u$ and $q_2=d$ for the proton interpolating
fields, each basis can be expressed as follows:
\begin{align}
\epsilon_{abc}[u^T_aC\gamma_\mu u_b]\gamma_5\gamma^\mu
d_c&=4\epsilon_{abc} \left( [u^T_{R,a}C d_{R,b}]u_{L,c}-[u^T_{L,a}C
d_{L,b}]u_{R,c}
\right),\label{ncrnt1}\\
\epsilon_{abc}[u^T_aC\sigma_{\mu\nu} u_b]\gamma_5 \sigma^{\mu\nu}
d_c&=4\epsilon_{abc} \left( [u^T_{R,a}C d_{R,b}]u_{R,c}-[u^T_{L,a}C
d_{L,b}]u_{L,c} \right)\label{ncrnt2},
\end{align}
where the basis \eqref{ncrnt1} is known as Ioffe's interpolating
fields for nucleons.

In the vacuum, the lowest
dimensional quark operator appearing in the OPE of the self-correlation function of basis \eqref{ncrnt2}, which can also be obtained from taking $t=1$ in Eq.~\eqref{ncrnt},    starts from the dimension-9 six-quark condensate term, whose Wilson coefficient comes from
perturbative gluon attachment for the external momentum flow with all quarks disconnected.   On the other hand,  for the  self-correlation function of the basis \eqref{ncrnt1}, obtained by taking $t=-1$, the OPE starts from the dimension-3
chiral condensate.
Hence, as
we do not know the exact value of the six-quark condensate and the
$\alpha_s$ is sensitive to subtraction scale ($\sim
1~\textrm{GeV}$), a reliable OPE can be obtained in the vacuum by choosing  the  basis \eqref{ncrnt1} (Ioffe's choice).
In the medium, the dimension-3 quark density operator appears in the OPE for both currents.  However, we retain the vacuum choice for the basis because we want to build the correction from the medium starting from a reliable vacuum OPE.
 As in the case of $\Sigma$, the diquark structure
in Ioffe's choice \eqref{ncrnt1} is a pseudovector $q^T
C\gamma_\mu q $ ($s^P=0^{-}$ for $\mu=0$, $s^P=1^{+}$ for $\mu=i$),
where $q$ is $u$ for the proton and $d$ for the neutron.

It is generally believed that the most attractive diquark is the
scalar channel $\epsilon_{abc}[ u^T_a C \gamma_5 d_b]$. However, because
this channel is in the iso-spin asymmetric combination, it can not
occur in the $\Sigma$ channel.  Also, in the proton channel, the
$ud$ diquark could be either in $I=0$ or $I=1$. This fact is in
stark contrast to the $\Lambda$ case, where its $ud$ quantum number
can be identified with the most attractive channel.  As we will see,
sum-rule analysis indeed favors the dominance of the most attractive
diquark channel for the $\Lambda$ interpolating fields.

\subsection{Interpolating fields for the $\Lambda$ hyperon}
For the $\Lambda$ hyperon, the basis set can be taken just for the
set~\eqref{di-asym} with $q_3=s$.  One can try another approach by
requiring the following conditions: (i) the diquarks in the $s=0$ configuration and
(ii) the light quarks in $I=0$ configuration. The following basis
set can then be considered:
\begin{align}
\left\{\epsilon_{abc}[u^T_a C d_b]\gamma_5 s_c,~\epsilon_{abc}[u^T_a
C \gamma_5 d_b] s_c,~\epsilon_{abc}\left([u^T_a C s_b]\gamma_5
d_c-[d^T_a C s_b]\gamma_5 u_c \right),~\epsilon_{abc}\left([u^T_a
C\gamma_5  s_b] d_c-[d^T_a C\gamma_5  s_b] u_c \right)
\right\}.\label{lset}
\end{align}
Using the Fierz rearrangement, the third and fourth basis can be
rearranged as
\begin{align}
\epsilon_{abc}\left([u^T_a C s_b]\gamma_5 d_c-[d^T_a C s_b]\gamma_5
u_c \right)&= \frac{1}{2}\epsilon_{abc}\left([u^T_aC d_b]\gamma_5
s_c+[u^T_a C \gamma_5 d_b] s_c-[u^T_a C\gamma_5\gamma_\mu
d_b]\gamma^\mu s_c\right),\label{lfrz1}\\
\epsilon_{abc}\left([u^T_a C\gamma_5  s_b] d_c-[d^T_a C\gamma_5 s_b]
u_c \right)&=\frac{1}{2}\epsilon_{abc}\left([u^T_aC d_b]\gamma_5
s_c+[u^T_a C \gamma_5 d_b] s_c+[u^T_a C\gamma_5\gamma_\mu
d_b]\gamma^\mu s_c \right).\label{lfrz2}
\end{align}
Hence, the basis set \eqref{lset} can be reduced to the
set~\eqref{di-asym} with $q_3=s$:
\begin{align}
\{\epsilon_{abc}[u^T_a C d_b]\gamma_5 s_c,~\epsilon_{abc}[u^T_a C
\gamma_5 d_b] s_c,~\epsilon_{abc}[u^T_a C\gamma_5\gamma_\mu
d_b]\gamma^\mu s_c \}.\label{lset2}
\end{align}
The generalized interpolating fields can then be written as follows:
\begin{align}
\eta_{\Lambda(\tilde{a},\tilde{b})}&=A_{(\tilde{a},\tilde{b})}\epsilon_{abc}\left([u^T_aC
d_b]\gamma_5 s_c+\tilde{a}[u^T_a C \gamma_5 d_b] s_c+\tilde{b}[u^T_a
C\gamma_5\gamma_\mu d_b]\gamma^\mu s_c\right),\label{lcrnt}
\end{align}
where $A_{(\tilde{a},\tilde{b})}$ is an overall normalization
constant. As the self-energies will be obtained by taking the ratios
of Borel transformed invariants, the overall normalization becomes
irrelevant and the free parameters can be reduced to $\tilde{a}$ and
$\tilde{b}$.  The basis set can be written in terms of helicity
states:
\begin{align}
\epsilon_{abc}[u^T_aC d_b]\gamma_5 s_c&=\epsilon_{abc}\left( [u^T_{R,a}C d_{R,b}]s_{R,c}-[u^T_{R,a}C d_{R,b}]s_{L,c} - (L \leftrightarrow R) \right), \label{asymdic1}\\
\epsilon_{abc}[u^T_aC\gamma_5 d_b] s_c&=\epsilon_{abc}\left( [u^T_{R,a}C d_{R,b}]s_{R,c}+[u^T_{R,a}C d_{R,b}]s_{L,c} - (L \leftrightarrow R) \right), \label{asymdic2}\\
\epsilon_{abc}[u^T_a C\gamma_5\gamma_\mu d_b]\gamma^\mu s_c &=
2\epsilon_{abc} \left( [u^T_{R,a}C s_{R,b}]d_{L,c}-[d^T_{R,a}C
s_{R,b}]u_{L,c} - (L \leftrightarrow R) \right). \label{asymdic3}
\end{align}
If one changes the $I=0$ combination on the right-hand side of basis
\eqref{asymdic3} to the $I=1$ combination, the basis changes into
Ioffe's choice for the $\Sigma$ family \eqref{scrnt1}. The light quark
condensate only appears in the cross correlation function between
bases \eqref{asymdic2} and \eqref{asymdic3}. The lowest dimensional
quark condensates in the self-correlation function of each basis is
$\langle \bar{s} s \rangle $.  Hence, determining the parameters in
Eq.~\eqref{lcrnt} corresponds to determining the weight of lowest
dimensional operators $\langle \bar{q} q \rangle $ and $\langle
\bar{s} s \rangle $ in the OPE of the correlation function.

The commonly used interpolating fields for $\Lambda$, called Ioffe's
choice, can be obtained by choosing
$\{\tilde{a},\tilde{b}\}=\{-1,-1/2\}$ in Eq.~\eqref{lcrnt}:
\begin{align}
\eta_{\Lambda(-1,-1/2)}&\Rightarrow\sqrt{\frac{2}{3}}\epsilon_{abc}\left([u^T_a
C\gamma_\mu s_b]\gamma_5\gamma^\mu d_c-[d^T_aC\gamma_\mu
s_b]\gamma_5\gamma^\mu u_c\right),\label{Ilcrnt}
\end{align}
where the overall normalization $A_{(-1,-1/2)}=\sqrt{8/3}$ has been
obtained from SU(3) flavor transformation from Ioffe's interpolating
field for nucleon \eqref{ncrnt1}. The choice for $\tilde{a}=-1$
causes large cancellation between the OPE terms in the scalar
invariants. The canceled part comes from the self-correlation
function of basis \eqref{asymdic1} and of basis \eqref{asymdic2}.
This means that, in studies where the Ioffe's choice \eqref{Ilcrnt} were
used, a large portion of the scalar invariant
$\Pi_s(q_0,\vert\vec{q}\vert)$ comes from the self-correlation
function of basis \eqref{asymdic3}. In this choice, as can be found
in next section, the chiral condensate term has a larger weight than
the strange quark condensate and the perturbative contributions,
making the OPE less reliable. Moreover, as can be seen in the next
section, phenomenological implications strongly suggest that taking
a large $\tilde{a} $ and a small $\tilde{b}$ value  gives a most
efficient sum rule and hence the best choice for the interpolating
fields. The argument for determining the stable region in
$\{\tilde{a},\tilde{b}\}$ plane will be given in the next sections.

The  diquark structure in the bases \eqref{asymdic1},
\eqref{asymdic2}, and \eqref{asymdic3} are pseudoscalar $u^T C d$
($s^P=0^{-}$), scalar $u^T C\gamma_5 d$ ($s^P=0^{+}$) and vector
$u^T C\gamma_5\gamma_\mu d$ ($s^P=0^{+}$ for $\mu=0$, $s^P=1^{-}$
for $\mu=i$) respectively. The corresponding relative angular
momentum between the light quarks in the nonrelativistic
limit are
$l=1$ for pseudoscalar and $l=0$ for scalar and vector diquark.

\section{Operator Product Expansion and Borel sum rules}\label{sec3}

In this section, we will list the OPE of the generalized $\Lambda$
correlation function and only the explicit four-quark OPE terms of
$\Sigma^{+}$ correlation function. The other OPE terms of the
nucleon and $\Sigma^{+}$ correlation function with Ioffe's choice
can be found in Refs.~\cite{Jeong:2012pa, Jin:1994bh}. Also,
covariant derivative expansion and factorization hypothesis have
been minimally used.

\subsection{Brief summary for the condensates and input parameters}\allowdisplaybreaks

A detailed description for the in-medium light quark and gluon condensates can be found in Refs.~\cite{Cohen:1991nk, Furnstahl:1992pi, Jin:1992id, Jin:1993up, Jeong:2012pa}. The strange quark condensate can be written as follows:
\begin{align}
\langle \bar{s}s \rangle_{\rho,I}= \langle \bar{s}s
\rangle_{\textrm{vac}}+\langle \bar{s}s \rangle_p \rho,
\end{align}
where $\langle \bar{s}s \rangle_{\textrm{vac}} $ is taken to be $
(0.8)\langle \bar{q}q \rangle_{\textrm{vac}}$ \cite{Shifman:1978bx,
Reinders:1984sr}, whereas the medium part can be determined by
constraining  the parameter $y=\langle \bar{s}s \rangle_p / \langle
\bar{q}q \rangle_p = m_q\sigma_{sN}/m_s\sigma_N $, which represents the
strange quark content in the proton. Recent theoretical developments including chiral effective theory~\cite{Alarcon:2012nr} and lattice QCD studies~\cite{Ren:2014vea, Durr:2015dna, Yang:2015uis} confine $y \leq
0.2$. We will take $y= 0.1$ throughout
this work.
For nonstrange nuclear matter, $\langle s^\dagger s \rangle_{\rho,I}=0$, whereas $\langle s^\dagger s \rangle_{\rho,I}$ will be nonzero if hyper-nuclear matter appears at the high density
regime. The covariant derivative expansions of the strange quark condensates can be written as
\begin{align}
\langle s^\dagger iD_0 s  \rangle_{\rho,I}& = \langle s^\dagger iD_0
s
 \rangle_{\textrm{vac}} +\langle s^\dagger iD_0 s  \rangle_p \rho
= \frac{m_s}{4}\langle \bar{s}s \rangle_{\textrm{vac}} + \left (
\frac{m_s}{4}\langle \bar{s}s \rangle_{p}+ \frac{3}{8}M_pA^s_2
\right)\rho,
\end{align}
where $A^s_2
=0.050$ \cite{Jin:1993fr}.

In this paper, we have changed some  definition of the symbols for the four-quark condensates as compared with Refs.~\cite{Choi:1993cu, Jeong:2012pa}. The new definition can be written as
\begin{align}
\epsilon_{abc}\epsilon_{a'b'c}\langle \bar{q}_{1a'} \Gamma_m^\alpha
q_{1a} \bar{q}_{2b'} \Gamma_m^{\beta}
q_{2b}\rangle_{\rho,I}=&~\frac{1}{4}g^{\alpha\beta}\langle
\bar{q}_{1} \Gamma_m  q_{1} \bar{q}_{2} \Gamma_m
q_{2}\rangle_{\textrm{tr.}}+\left(u^{\alpha}u^\beta
-\frac{1}{4}g^{\alpha\beta}\right)\langle \bar{q}_{1} \Gamma_m q_{1}
\bar{q}_{2} \Gamma_m
q_{2}\rangle_{\textrm{s.t.}},\\
\langle \bar{q}_{1} \Gamma_m q_{1} \bar{q}_{2} \Gamma_m
q_{2}\rangle_{\textrm{tr.}} =&~\frac{2}{3}\langle\bar{q}_1
\Gamma_m^\alpha q_1 \bar{q}_2 \Gamma_{m\alpha}
q_2\rangle_{\textrm{vac}}-2\langle\bar{q}_1 \Gamma_m^\alpha t^A q_1
\bar{q}_2 \Gamma_{m\alpha} t^A
q_2\rangle_{\textrm{vac}}\nonumber\\
& +\sum_{i=\{n,p\}}\left(\frac{2}{3}\langle\bar{q}_1 \Gamma_m^\alpha
q_1 \bar{q}_2 \Gamma_{m\alpha} q_2\rangle_{i}-2\langle\bar{q}_1
\Gamma_m^\alpha t^A q_1 \bar{q}_2 \Gamma_{m\alpha} t^A
q_2\rangle_{i}\right)\rho_i,\\
\langle \bar{q}_{1} \Gamma_m q_{1} \bar{q}_{2} \Gamma_m
q_{2}\rangle_{ \textrm{s.t.}}
=&~\sum_{i=\{n,p\}}\left(\frac{2}{3}\langle\bar{q}_1 \Gamma_m q_1
\bar{q}_2 \Gamma_m q_2\rangle_{i, \textrm{s.t.}} -2\langle\bar{q}_1
\Gamma_m t^A q_1 \bar{q}_2 \Gamma_n t^A
q_2\rangle_{i,\textrm{s.t.}}\right)\rho_i,
\end{align}
where $q_1$, $q_2$ represent the quark flavor,
$\Gamma_m=\{I,\gamma_5, \gamma, \gamma_5\gamma, \sigma \}$ and each
subscript $ \textrm{vac} $, $ i $, and $ \textrm{s.t.} $ represents
the vacuum expectation value, nucleon expectation value, and symmetric
traceless matrix element, respectively. Twist-4 matrix elements
$\langle\bar{q}_1 \Gamma_m  q_1 \bar{q}_2 \Gamma_m q_2\rangle_{i,
\textrm{s.t.}}$ and $\langle\bar{q}_1 \Gamma_m t^A q_1 \bar{q}_2
\Gamma_m  t^A q_2\rangle_{i,\textrm{s.t.}}$ can be estimated from
DIS data following the arguments presented in
Refs.~\cite{Choi:1993cu, Jeong:2012pa}. For spin-0 and spin-1
operators, the factorization hypothesis has been used:
\begin{align}
\langle q^a_\alpha\bar{q}^b_\beta
q^c_\gamma\bar{q}^d_\delta\rangle_{\rho,I}&\simeq\langle
q^a_\alpha\bar{q}^b_\beta \rangle_{\rho,I}\langle
q^c_\gamma\bar{q}^d_\delta \rangle_{\rho,I}- \langle
q^a_\alpha\bar{q}^d_\delta \rangle_{\rho,I} \langle
q^c_\gamma\bar{q}^b_\beta \rangle_{\rho,I},\nonumber\\
 \langle q1^a_{1\alpha} \bar{q}^b_{1\beta}
 q^c_{2\gamma} \bar{q}^d_{2\delta}
\rangle_{\rho,I}& \simeq \langle q^a_{1\alpha} \bar{q}^b_{1\beta}
\rangle_{\rho,I}  \langle q^c_{2\gamma} \bar{q}^d_{2\delta}
\rangle_{\rho,I}.\label{factorization}
\end{align}
This factorization scheme  can only be justified in the vacuum and in the large-$N_c$ limit.  In general, model calculations find large violations depending on the stucture of the four-quark operator~\cite{GomezNicola:2010tb, Drukarev:2012av}. Therefore, we will introduce the following parametrized form for the medium dependence of the four-quark operators and investigate the result as a function of the parameters that will probe different factorized forms and values for the vacuum and medium part of the four-quark operators separately.   After taking the average of color and Dirac index, the remaining scalar
condensates have been parametrized as
\begin{align}
\langle[\bar{q}q]_{u,d}\rangle^2_{\rho,I } & \Rightarrow
k_1\langle\bar{q}q\rangle^2_{\textrm{vac}} +
2f_1\left(\langle[\bar{q}q]_0 \rangle_p \mp\langle[\bar{q}q]_1
\rangle_p I\right)\langle\bar{q}q\rangle_{\textrm{vac}}
\rho,\label{4qscalar}\\
\langle \bar{u}u \rangle_{\rho,I} \langle \bar{d}d\rangle_{\rho,I}&
\Rightarrow k_1 \langle\bar{q}q\rangle^2_{\textrm{vac}} + 2f_1
\langle[\bar{q}q]_0
\rangle_p  \langle\bar{q}q\rangle_{\textrm{vac}}\rho,\label{4qscalar1}\\
\langle \bar{q}q \rangle_{\rho,I} \langle \bar{s}s\rangle_{\rho,I}&
\Rightarrow
k_2\langle\bar{q}q\rangle_{\textrm{vac}}\langle\bar{s}s\rangle_{\textrm{vac}}
+ f_2\left( \langle\bar{s}s \rangle_p
\langle\bar{q}q\rangle_{\textrm{vac}}+\langle[\bar{q}q]_0 \rangle_p
\langle\bar{s}s\rangle_{\textrm{vac}}+\langle[\bar{q}q]_1 \rangle_p
\langle\bar{s}s\rangle_{\textrm{vac}}I \right)\rho,\label{4qscalar2}
\end{align}
where parameters $k_1$, $k_2$ determine the vacuum strength and $f_1$,
$f_2$ determine the medium dependence of the scalar four-quark
condensate. Both $k_1$ and $k_2$ are set to be 1 as the in-medium sum
rules do not show drastic change in the range $0.5 \leq k_1,k_2 \leq
3.0$ as one can find in Appendix~\ref{appentw}. According to
previously reported studies, $\vert f_1 \vert $ should be weak
($\vert f_1 \vert \ll 1$)
 \cite{Cohen:1994wm, Jeong:2012pa} but $f_2$ can be strong ($f_2 \simeq 1$). This scale
difference between the condensates seems reasonable because the
strange quark operator only has sea quark contributions in the normal
nuclear matter expectation value whereas the light quark operator
has additional valence quark contributions. Detailed arguments for
the twist-4 matrix elements and the parameter dependencies are
presented in Appendix~\ref{appentw}.

As each quasinucleon has its own quasi-Fermi sea, the
external three-momentum of the quasibaryon will be set at the  Fermi
 momentum at the given nuclear matter density:
$\vert\vec{q}\vert=270~\textrm{MeV}$ when $\rho=\rho_0=
0.16~\textrm{fm}^{-3}=(110~\textrm{MeV})^3$. For the same reason,
the external momentum for the quasihyperon will be set to
$\vert\vec{q}\vert=0~\textrm{MeV}$.

The correlation function contains all possible
resonances that overlap with the quantum number of the interpolating
fields as discussed before. As our interests are the self-energies on the quasibaryon
pole, the other excitations should be suppressed. Borel sum rules
can be used for this purpose: the weight function $W(\omega)=
(\omega-\bar{E}_q)e^{-\omega^2/M^2}$ has been applied to the
discontinuity in the dispersion relation \eqref{dis-eo} and the
corresponding differential operator $\bar{\mathcal{B}}$ has been
applied to the OPE side. Each transformed part will be denoted as
$\overline{\mathcal{W}}_M[\Pi(q_0^2,\vert\vec{q}\vert)]$ and
$\bar{\mathcal{B}}[\Pi(q_0^2,\vert\vec{q}\vert)]$ respectively.
Details for weighting scheme and corresponding differential operation in the Borel sum rules that we use in this work can be found in Refs.~\cite{Furnstahl:1992pi, Cohen:1994wm, Beneke:1994rs, Beneke:1994sw}.

Borel transformed invariants contain the quasi-antipole $\bar{E}_q$
as an input parameter. As we are following relativistic-mean-field-type phenomenology, the antipole $\bar{E}_q$ is already defined
regardless of the actual existence of the pole in the medium.  The exact
value can be determined by solving the self-consisted dispersion
relation:
\begin{align}
\bar{E}_q=\Sigma_v(\bar{E}_q)-\sqrt{\vec{q}^2+M^{*}(\bar{E}_q)^{2}}.\label{qhd}
\end{align}
The exact solution of this relation has been used for the antipole
value. In density plot, we consider two choices. First, as the
quasi-antibaryon excitation may be broadened and may not exist as
a pole in the nuclear matter, the value can just be taken as a
constant calculated at the saturation
nuclear matter density. Second,
one can calculate the exact solution of Eq.~\eqref{qhd} self
consistently at a given density. Because the condensates are
approximated with linear density approximation, the sum rule itself
and the solution of Eq.~\eqref{qhd} are expected to be valid up to
densities slightly above the saturation density.

\subsection{OPE of the generalized $\Lambda$ correlation function}

The OPE of the generalized $\Lambda$ correlation function can be
calculated as follows:
\begin{align}
\Pi^e_{\Lambda,s}(q_0^2,\vert\vec{q}\vert) =&~
\frac{(1-\tilde{a}^2+2\tilde{b}^2)}{128\pi^4}m_s(q^2)^2\ln(-q^2)\nonumber\\
&-\frac{(1-\tilde{a}^2+2\tilde{b}^2)}{16\pi^2}q^2\ln(-q^2)\langle
\bar{s}s \rangle_{\rho,I}
+\frac{\tilde{a}\tilde{b}}{4\pi^2}q^2\ln(-q^2)\langle \bar{q}q
\rangle_{\rho,I}\nonumber\\
&+\frac{(1-\tilde{a}^2-2\tilde{b}^2)}{128\pi^2}m_s\ln(-q^2)\left\langle\frac{\alpha_s}{\pi}G^2\right\rangle_{\rho,I}\nonumber\\
&-\frac{(1+\tilde{a}^2+4\tilde{b}^2)}{4}\frac{m_s}{q^2}\langle
\bar{u}u\bar{d}d
\rangle_{\textrm{tr.}}-\frac{(1+\tilde{a}^2-4\tilde{b}^2)}{4}\frac{m_s}{q^2}\langle
\bar{u} \gamma_5 u\bar{d} \gamma_5
d\rangle_{\textrm{tr.}}+\frac{(1-\tilde{a}^2-2\tilde{b}^2)}{4}\frac{m_s}{q^2}\langle
\bar{u} \gamma  u\bar{d} \gamma  d
\rangle_{\textrm{tr.}}\nonumber\\
&+\frac{(1-\tilde{a}^2+2\tilde{b}^2)}{4}\frac{m_s}{q^2}\langle
\bar{u} \gamma_5\gamma  u\bar{d} \gamma_5\gamma  d
\rangle_{\textrm{tr.}}+\frac{(1+\tilde{a}^2)}{8}\frac{m_s}{q^2}\langle
\bar{u}
\sigma  u\bar{d} \sigma  d \rangle_{\textrm{tr.}},\\
\Pi^o_{\Lambda,s}(q_0^2,\vert\vec{q}\vert) =&-
\frac{(1-\tilde{a}^2+2\tilde{b}^2)}{8\pi^2}m_s\ln(-q^2) \langle
q^\dagger
q\rangle_{\rho,I}-\frac{2\tilde{a}\tilde{b}}{3}\frac{1}{q^2}\langle
q^\dagger q \rangle_{\rho,I} \langle \bar{q}  q
\rangle_{\textrm{vac}}+\frac{(1-\tilde{a}^2+4\tilde{b}^2)}{3}\frac{1}{q^2}\langle
q^\dagger q \rangle_{\rho,I} \langle \bar{s}
s\rangle_{\textrm{vac}},\\
\Pi^e_{\Lambda,q}(q_0^2,\vert\vec{q}\vert) =& - \frac{
(1+\tilde{a}^2+4\tilde{b}^2)}{512\pi^4}(q^2)^2\ln(-q^2)+\frac{\tilde{a}\tilde{b}}{4\pi^2}m_s\ln(-q^2)\langle
\bar{q} q
\rangle_{\rho,I}-\frac{(1+\tilde{a}^2+4\tilde{b}^2)}{32\pi^2}m_s\ln(-q^2)\langle
\bar{s} s
\rangle_{\rho,I}\nonumber\\
&-\frac{(1+\tilde{a}^2+4\tilde{b}^2)}
{256\pi^2}\ln(-q^2)\left\langle\frac{\alpha_s}{\pi}G^2\right\rangle_{\rho,I}
\nonumber\\
&+\frac{(1-\tilde{a}^2+2\tilde{b}^2)}{4}\frac{1}{q^2}\langle
\bar{u}u\bar{d}d
\rangle_{\textrm{tr.}}+\frac{(1-\tilde{a}^2-2\tilde{b}^2)}{4}\frac{1}{q^2}\langle
\bar{u} \gamma_5 u\bar{d} \gamma_5
d\rangle_{\textrm{tr.}}-\frac{(1+\tilde{a}^2-\tilde{b}^2)}{4}\frac{1}{q^2}\langle
\bar{u} \gamma  u\bar{d} \gamma d
\rangle_{\textrm{tr.}}\nonumber\\
&-\frac{(1+\tilde{a}^2+\tilde{b}^2)}{4}\frac{1}{q^2}\langle \bar{u}
\gamma_5\gamma  u\bar{d} \gamma_5\gamma d
\rangle_{\textrm{tr.}}-\frac{(1-\tilde{a}^2)}{8}\frac{1}{q^2}\langle
\bar{u}
\sigma  u\bar{d} \sigma  d \rangle_{\textrm{tr.}}\nonumber\\
&- \tilde{a}\tilde{b}\frac{1}{q^2}\langle \bar{q}q\bar{s}s
\rangle_{\textrm{tr.}}- \tilde{b}\frac{1}{q^2}\langle \bar{q}
\gamma_5 q\bar{s} \gamma_5
s\rangle_{\textrm{tr.}}-\frac{(1+\tilde{a}^2-10\tilde{b}^2)}{8}\frac{1}{q^2}\langle
\bar{q} \gamma  q\bar{s} \gamma  s
\rangle_{\textrm{tr.}}\nonumber\\
&-\frac{(\tilde{a}-3\tilde{b}^2)}{4}\frac{1}{q^2}\langle \bar{q}
\gamma_5\gamma  q \bar{s} \gamma_5\gamma s
\rangle_{\textrm{tr.}}+\frac{\tilde{a}\tilde{b}}{4}\frac{1}{q^2}\langle
\bar{q} \sigma  q\bar{s} \sigma  s
\rangle_{\textrm{tr.}}\nonumber\\
&+\frac{\tilde{b}^2}{4}\frac{1}{q^2}\langle \bar{u} \gamma u\bar{d}
\gamma d
\rangle_{\textrm{s.t.}}-\frac{\tilde{b}^2}{4}\frac{1}{q^2}\langle
\bar{u} \gamma_5\gamma u\bar{d} \gamma_5\gamma d
\rangle_{\textrm{s.t.}}+\frac{\tilde{b}^2}{4}\frac{1}{q^2}\langle
\bar{u} \sigma u\bar{d} \sigma d
\rangle_{\textrm{s.t.}}\nonumber\\
&+\frac{(1+\tilde{a}^2-2\tilde{b}^2)}{8}\frac{1}{q^2}\langle \bar{q}
 \gamma q\bar{s} \gamma s
\rangle_{\textrm{s.t.}}+\frac{(\tilde{a}+\tilde{b}^2)}{4}\frac{1}{q^2}\langle
\bar{q} \gamma_5\gamma q\bar{s} \gamma_5\gamma s
\rangle_{\textrm{s.t.}}
-\frac{\tilde{a}\tilde{b}}{4}\frac{1}{q^2}\langle \bar{q} \sigma
q\bar{s} \sigma \gamma  s
\rangle_{\textrm{s.t.}}, \\
\Pi_{\Lambda,q}^o(q_0^2,\vert\vec{q}\vert) =&~
\frac{(1+\tilde{a}^2+2\tilde{b}^2)}{24\pi^2}\ln(-q^2)\langle
q^\dagger q\rangle_{\rho,I},\\
\Pi_{\Lambda,u}^e(q_0^2,\vert\vec{q}\vert)=&~\frac{(1+\tilde{a}^2+14\tilde{b}^2)}{48\pi^2}q^2\ln(-q^2)\langle
q^\dagger q \rangle_{\rho,I} -\frac{2\tilde{a}\tilde{b}}{3}
\frac{m_s}{q^2}\langle q^\dagger q
\rangle_{\rho,I} \langle \bar{s} s\rangle_{\textrm{vac}}, \\
\Pi_{\Lambda,u}^o(q_0^2,\vert\vec{q}\vert)=&-
\tilde{b}^2\frac{1}{q^2}\langle \bar{u} \gamma u\bar{d} \gamma d
\rangle_{\textrm{s.t.}}+\tilde{b}^2\frac{1}{q^2}\langle \bar{u}
\gamma_5\gamma u\bar{d} \gamma_5\gamma d
\rangle_{\textrm{s.t.}}-\tilde{b}^2\frac{1}{q^2}\langle \bar{u}
\sigma u\bar{d} \sigma d
\rangle_{\textrm{s.t.}}\nonumber\\
&-\frac{(1+\tilde{a}^2-2\tilde{b}^2)}{2} \frac{1}{q^2}\langle
\bar{u} \gamma u\bar{d} \gamma d
\rangle_{\textrm{s.t.}}-(\tilde{a}+\tilde{b}^2) \frac{1}{q^2}\langle
\bar{u} \gamma_5\gamma u\bar{d} \gamma_5\gamma d
\rangle_{\textrm{s.t.}}+\tilde{a}\tilde{b}\frac{1}{q^2}\langle
\bar{u} \sigma u\bar{d} \sigma d \rangle_{\textrm{s.t.}} ,
\end{align}
where the normalization constant is chosen to be
$A_{(\tilde{a},\tilde{b})}\rightarrow 1$ and the spin-1 four-quark
condensates are listed in the factorized forms. The operator in the light quark flavor $q$ is defined as $\bar{q}\Gamma q \equiv (\bar{u}\Gamma u+\bar{d}\Gamma d)/2$. Borel transformed
invariants can be summarized as
\begin{align}
\overline{\mathcal{W}}_M^{\textrm{subt.}}[\Pi_{\Lambda,s}(q_0^2,\vert\vec{q}\vert)]
= &~
\lambda_\Lambda^{{*}2}M_\Lambda^{*}e^{-(E_{\Lambda,q}^2-\vec{q}^2)/M^2}= \bar{\mathcal{B}}[\Pi^e_{\Lambda,s}(q_0^2,\vert\vec{q}\vert)]_{\textrm{subt.}}-\bar{E}_{\Lambda,q}\bar{\mathcal{B}}[\Pi^o_{\Lambda,s}(q_0^2,\vert\vec{q}\vert)]_{\textrm{subt.}}\nonumber\\
=&-\frac{(1-\tilde{a}^2+2\tilde{b}^2)}{64\pi^4}m_s (M^2)^3 \tilde{E}_2  L^{-\frac{8}{9}} \nonumber\\
&+\frac{(1-\tilde{a}^2+2\tilde{b}^2)}{16\pi^2}(M^2)^2\langle
\bar{s}s \rangle_{\rho,I} \tilde{E}_1
-\frac{\tilde{a}\tilde{b}}{4\pi^2} (M^2)^2 \langle \bar{q}q
\rangle_{\rho,I} \tilde{E}_1\nonumber\\
&-\frac{(1-\tilde{a}^2-2\tilde{b}^2)}{128\pi^2}m_s M^2 \left\langle\frac{\alpha_s}{\pi}G^2\right\rangle_{\rho,I} \tilde{E}_0 L^{-\frac{8}{9}}\nonumber\\
&+\frac{(1+\tilde{a}^2+4\tilde{b}^2)}{4}m_s \langle \bar{u}u\bar{d}d
\rangle_{\textrm{tr.}} +\frac{(1+\tilde{a}^2-4\tilde{b}^2)}{4}m_s
\langle \bar{u} \gamma_5 u\bar{d} \gamma_5
d\rangle_{\textrm{tr.}}\nonumber\\
&-\frac{(1-\tilde{a}^2-2\tilde{b}^2)}{4}m_s \langle \bar{u} \gamma
u\bar{d} \gamma  d
\rangle_{\textrm{tr.}}-\frac{(1-\tilde{a}^2+2\tilde{b}^2)}{4}m_s
\langle \bar{u} \gamma_5\gamma  u\bar{d} \gamma_5\gamma  d
\rangle_{\textrm{tr.}}- \frac{(1+\tilde{a}^2)}{8}m_s \langle \bar{u}
\sigma  u\bar{d} \sigma d
\rangle_{\textrm{tr.}}\nonumber\\
&-\bar{E}_{\Lambda,q} \Bigg[\frac{(1-\tilde{a}^2+
2\tilde{b}^2)}{8\pi^2} m_s M^2 \langle q^\dagger q\rangle_{\rho,I}
\tilde{E}_0 L^{-\frac{8}{9}}\nonumber\\
&\qquad\qquad+\frac{2 \tilde{a}\tilde{b}}{3} \langle q^\dagger q
\rangle_{\rho,I} \langle \bar{q}  q
\rangle_{\textrm{vac}}-\frac{(1-\tilde{a}^2+ 4\tilde{b}^2)}{3}
\langle q^\dagger q \rangle_{\rho,I} \langle \bar{s}
s\rangle_{\textrm{vac}}\Bigg],\\
\overline{\mathcal{W}}_M^{\textrm{subt.}}[\Pi_{\Lambda,
q}(q_0^2,\vert\vec{q}\vert)] = &~
\lambda_\Lambda^{{*}2} e^{-(E_{\Lambda,q}^2-\vec{q}^2)/M^2}= \bar{\mathcal{B}}[\Pi^e_{\Lambda,q}(q_0^2,\vert\vec{q}\vert)]_{\textrm{subt.}}-\bar{E}_{\Lambda,q}\bar{\mathcal{B}}[\Pi^o_{\Lambda,q}(q_0^2,\vert\vec{q}\vert)]_{\textrm{subt.}}\nonumber\\
=& ~ \frac{ (1+\tilde{a}^2+4\tilde{b}^2)}{256\pi^4}(M^2)^3
\tilde{E}_2 L^{-\frac{4}{9}}+\frac{(1+\tilde{a}^2+4\tilde{b}^2)}
{256\pi^2} M^2
\left\langle\frac{\alpha_s}{\pi}G^2\right\rangle_{\rho,I}
\tilde{E}_0
L^{-\frac{4}{9}}\nonumber\\
&-\frac{\tilde{a}\tilde{b}}{4\pi^2}m_s M^2\langle \bar{q} q
\rangle_{\rho,I} \tilde{E}_0
L^{-\frac{4}{9}}+\frac{(1+\tilde{a}^2+4\tilde{b}^2)}{32\pi^2}m_s M^2
\langle \bar{s} s \rangle_{\rho,I} \tilde{E}_0 L^{-\frac{4}{9}}
\nonumber\\
&-\frac{(1-\tilde{a}^2+2\tilde{b}^2)}{4} \langle \bar{u}u\bar{d}d
\rangle_{\textrm{tr.}} -\frac{(1-\tilde{a}^2-2\tilde{b}^2)}{4}
\langle \bar{u} \gamma_5 u\bar{d} \gamma_5
d\rangle_{\textrm{tr.}}+\frac{(1+\tilde{a}^2-\tilde{b}^2)}{4}
\langle \bar{u} \gamma  u\bar{d} \gamma  d
\rangle_{\textrm{tr.}}\nonumber\\
&+\frac{(1+\tilde{a}^2+\tilde{b}^2)}{4} \langle \bar{u}
\gamma_5\gamma  u\bar{d} \gamma_5\gamma  d
\rangle_{\textrm{tr.}}+\frac{(1-\tilde{a}^2)}{8} \langle \bar{u}
\sigma  u\bar{d} \sigma  d \rangle_{\textrm{tr.}}+
\tilde{a}\tilde{b} \langle \bar{q}q\bar{s}s \rangle_{\textrm{tr.}} +
\tilde{b} \langle \bar{q} \gamma_5 q\bar{s} \gamma_5
s\rangle_{\textrm{tr.}}\nonumber\\
&+\frac{(1+\tilde{a}^2-10\tilde{b}^2)}{8} \langle \bar{q} \gamma
q\bar{s} \gamma  s
\rangle_{\textrm{tr.}}+\frac{(\tilde{a}-3\tilde{b}^2)}{4} \langle
\bar{q} \gamma_5\gamma q \bar{s} \gamma_5\gamma  s
\rangle_{\textrm{tr.}}-\frac{\tilde{a}\tilde{b}}{4} \langle \bar{q}
\sigma  q\bar{s} \sigma  s
\rangle_{\textrm{tr.}}\nonumber\\
&-\frac{\tilde{b}^2}{4} \langle \bar{u} \gamma u\bar{d} \gamma d
\rangle_{\textrm{s.t.}}+\frac{\tilde{b}^2}{4} \langle \bar{u}
\gamma_5\gamma u\bar{d} \gamma_5\gamma d
\rangle_{\textrm{s.t.}}-\frac{\tilde{b}^2}{4} \langle \bar{u} \sigma
u\bar{d} \sigma d
\rangle_{\textrm{s.t.}}\nonumber\\
&-\frac{(1+\tilde{a}^2-2\tilde{b}^2)}{8} \langle \bar{q}
 \gamma q\bar{s} \gamma s
\rangle_{\textrm{s.t.}}+\frac{(\tilde{a}+\tilde{b}^2)}{4}\frac{1}{q^2}\langle
\bar{q} \gamma_5\gamma q\bar{s} \gamma_5\gamma s
\rangle_{\textrm{s.t.}}+\frac{\tilde{a}\tilde{b}}{4} \langle \bar{q}
\sigma q\bar{s} \sigma \gamma s
\rangle_{\textrm{s.t.}}\nonumber\\
&+\bar{E}_{\Lambda,q} \left[
\frac{(1+\tilde{a}^2+2\tilde{b}^2)}{24\pi^2} M^2 \langle
q^\dagger q\rangle_{\rho,I} \tilde{E}_0 \right],\\
\overline{\mathcal{W}}_M^{\textrm{subt.}}[\Pi_{\Lambda,u}(q_0^2,\vert\vec{q}\vert)]
= &~
\lambda_\Lambda^{{*}2} \Sigma^\Lambda_{v} e^{-(E_{\Lambda,q}^2-\vec{q}^2)/M^2}= \bar{\mathcal{B}}[\Pi^e_{\Lambda,u}(q_0^2,\vert\vec{q}\vert)]_{\textrm{subt.}}-\bar{E}_{\Lambda,q}\bar{\mathcal{B}}[\Pi^o_{\Lambda,u}(q_0^2,\vert\vec{q}\vert)]_{\textrm{subt.}}\nonumber\\
=&~\frac{(1+\tilde{a}^2+14\tilde{b}^2)}{48\pi^2}(M^2)^2\langle
q^\dagger q \rangle_{\rho,I}E_1
L^{-\frac{4}{9}}-\frac{2\tilde{a}\tilde{b}}{3}m_s \langle q^\dagger
q \rangle_{\rho,I} \langle \bar{s}
s\rangle_{\textrm{vac}} \nonumber\\
& + \bar{E}_{\Lambda,q} \Bigg[ \tilde{b}^2 \langle \bar{u} \gamma
u\bar{d} \gamma d \rangle_{\textrm{s.t.}}- \tilde{b}^2 \langle
\bar{u} \gamma_5\gamma u\bar{d} \gamma_5\gamma d
\rangle_{\textrm{s.t.}}+ \tilde{b}^2 \langle \bar{u} \sigma u\bar{d}
\sigma d
\rangle_{\textrm{s.t.}}\nonumber\\
&\qquad\qquad+\frac{(1+\tilde{a}^2-2\tilde{b}^2)}{2} \langle \bar{u}
\gamma u\bar{s} \gamma s
\rangle_{\textrm{s.t.}}+(\tilde{a}+\tilde{b}^2) \langle \bar{u}
\gamma_5\gamma u\bar{s} \gamma_5\gamma s
\rangle_{\textrm{s.t.}}-\tilde{a}\tilde{b} \langle \bar{u} \sigma
u\bar{s} \sigma s \rangle_{\textrm{s.t.}}\Bigg],
\end{align}
where $M$ is the Borel mass. The running corrections from the
anomalous dimensions are included as
\begin{align}
L^{-2\Gamma_\eta+\Gamma_{O_n}}\equiv\left[\frac{\ln(M/\Lambda_{\textrm{QCD}})}{\ln(\mu/\Lambda_{\textrm{QCD}})}\right]^{-2\Gamma_\eta+\Gamma_{O_n}},
\end{align}
where $\Gamma_\eta$ ($\Gamma_{O_n}$) is the anomalous dimension of
the interpolating fields $\eta$  ($\hat{O}_n$), and  $\mu$ is the
separation scale of the OPE taken to be $\mu\simeq 1~\textrm{GeV}$.
The continuum effect above ground resonance has been subtracted by
multiplying following $\tilde{E}_n$ to all $(M^2)^{n+1}$ terms in
$\overline{\mathcal{W}}_M[\Pi_{\Lambda,i}(q_0^2,\vert\vec{q}\vert)]
$ \cite{Furnstahl:1992pi, Cohen:1994wm}:
\begin{align}
\tilde{E}_0&\equiv1-e^{-s_0^{*}/M^2},\label{contsub0}\\
\tilde{E}_1&\equiv1-e^{-s_0^{*}/M^2}\left(s_0^{*}/M^2+1\right),\label{contsub1}\\
\tilde{E}_2&\equiv1-e^{-s_0^{*}/M^2}\left(s_0^{*2}/2M^4+s_0^{*}/M^2+1\right),\label{contsub2}
\end{align}
where $s_0^{*}\equiv\omega_0^{2}-\vec{q}^2$ and $\omega_0$ is the
energy at the continuum threshold taken as
$\omega_0=1.5~\textrm{GeV}$. The Borel transformed invariants after the continuum subtraction have been denoted as $\overline{\mathcal{W}}_M^{\textrm{subt.}}[\Pi_{\Lambda,i}(q_0^2,\vert\vec{q}\vert)]$.

\subsection{Four-quark condensate terms in the OPE of the  $\Sigma^{+}$ correlation function}
The OPE of the $\Sigma^{+}$ correlation function within Ioffe's
choice \eqref{scrnt1} is similar to the four-quark OPE of the
nucleon case. If one changes flavor $s \rightarrow d$ and neglects
$m_s$, the following OPE reduces to the nucleon case in
Ref.~\cite{Jeong:2012pa}. The four-quark-condensate terms in the OPE
of each invariant can be calculated as follows:
\begin{align}
\Pi_{\Sigma^{+}(4q),s}^e(q_0^2,\vert\vec{q}\vert)
=&-\frac{m_s}{q^2}\langle \bar{u} \gamma
u  \bar{u} \gamma u \rangle_{\textrm{tr.}} +\frac{m_s}{q^2}\langle \bar{u} \gamma_5 \gamma u  \bar{u}  \gamma_5 \gamma u \rangle_{\textrm{tr.}},\\
\Pi_{\Sigma^{+}(4q),s}^o(q_0^2,\vert\vec{q}\vert)
=&-\frac{4}{3}\frac{1}{q^2}\langle\bar{s}s\rangle_{\textrm{vac}}\langle
u^\dagger u \rangle_{\rho,I} ,\\
\Pi_{\Sigma^{+}(4q),q}^e(q_0^2,\vert\vec{q}\vert) =&-
\frac{1}{2q^2}\langle \bar{u} \gamma u  \bar{u} \gamma u
\rangle_{\textrm{tr.}}+ \frac{1}{2q^2}\langle \bar{u} \gamma_5
\gamma u \bar{u} \gamma_5 \gamma u
\rangle_{\textrm{tr.}}-\frac{5}{2q^2}\langle \bar{u}  \gamma u
\bar{s}   \gamma s \rangle_{\textrm{tr.}}-\frac{3}{2q^2}\langle
\bar{u} \gamma_5 \gamma u \bar{s} \gamma_5 \gamma s
\rangle_{\textrm{tr.}},\nonumber\\
&+ \frac{1}{2q^2}\langle \bar{u} \gamma u  \bar{u} \gamma u
\rangle_{\textrm{s.t.}}- \frac{1}{2q^2}\langle \bar{u} \gamma_5
\gamma u \bar{u} \gamma_5 \gamma u
\rangle_{\textrm{s.t.}}+\frac{1}{2q^2}\langle \bar{u}  \gamma u
\bar{s} \gamma s \rangle_{\textrm{s.t.}}-\frac{1}{2q^2}\langle
\bar{u} \gamma_5 \gamma u \bar{s} \gamma_5 \gamma s
\rangle_{\textrm{s.t.}},
\\
\Pi_{\Sigma^{+}(4q),u}^o(q_0^2,\vert\vec{q}\vert)=&-
\frac{2}{q^2}\langle \bar{u} \gamma u  \bar{u} \gamma u
\rangle_{\textrm{s.t.}}+ \frac{2}{q^2}\langle \bar{u} \gamma_5
\gamma u \bar{u} \gamma_5 \gamma u \rangle_{\textrm{s.t.}}-
\frac{2}{q^2}\langle \bar{u}  \gamma u \bar{s} \gamma s
\rangle_{\textrm{s.t.}}+\frac{2}{q^2}\langle \bar{u} \gamma_5 \gamma
u \bar{s} \gamma_5 \gamma s \rangle_{\textrm{s.t.}}.
\end{align}
Borel transformed invariants can be summarized as
\begin{align}
\overline{\mathcal{W}}_M^{\textrm{subt.}} [\Pi_{\Sigma^{+}(4q),s}(q_0^2,\vert\vec{q}\vert)]=&~ \bar{\mathcal{B}}[\Pi^e_{{\Sigma^{+}(4q)},s}(q_0^2,\vert\vec{q}\vert)]_{\textrm{subt.}}-\bar{E}_{\Sigma^{+},q}\bar{\mathcal{B}}[\Pi^o_{{\Sigma^{+}(4q)},s}(q_0^2,\vert\vec{q}\vert)]_{\textrm{subt.}}\nonumber\\
=&~m_s\langle \bar{u} \gamma u  \bar{u} \gamma u
\rangle_{\textrm{tr.}} - m_s\langle \bar{u} \gamma_5 \gamma u
\bar{u} \gamma_5 \gamma u
\rangle_{\textrm{tr.}}-\bar{E}_{\Sigma^{+},q}\frac{4}{3}\langle\bar{s}s\rangle_{\textrm{vac}}\langle
u^\dagger u \rangle_{\rho,I},\\
\overline{\mathcal{W}}_M^{\textrm{subt.}}[\Pi_{\Sigma^{+}(4q),q}
(q_0^2,\vert\vec{q}\vert)]
=&~\bar{\mathcal{B}}[\Pi^e_{{\Sigma^{+}(4q)},q}(q_0^2,\vert\vec{q}\vert)]_{\textrm{subt.}}-\bar{E}_{\Sigma^{+},q}\bar{\mathcal{B}}[\Pi^o_{{\Sigma^{+}(4q)},q}(q_0^2,\vert\vec{q}\vert)]_{\textrm{subt.}}\nonumber\\
=&~ \frac{1}{2}\langle \bar{u} \gamma u \bar{u} \gamma u
\rangle_{\textrm{tr.}}- \frac{1}{2}\langle \bar{u} \gamma_5 \gamma u
\bar{u} \gamma_5 \gamma u \rangle_{\textrm{tr.}}+\frac{5}{2}\langle
\bar{u} \gamma u \bar{s} \gamma s
\rangle_{\textrm{tr.}}+\frac{3}{2}\langle \bar{u} \gamma_5 \gamma u
\bar{s} \gamma_5 \gamma s
\rangle_{\textrm{tr.}}\nonumber\\
&- \frac{1}{2}\langle \bar{u} \gamma u  \bar{u} \gamma u
\rangle_{\textrm{s.t.}}+ \frac{1}{2}\langle \bar{u} \gamma_5 \gamma
u \bar{u} \gamma_5 \gamma u
\rangle_{\textrm{s.t.}}-\frac{1}{2}\langle \bar{u}  \gamma u \bar{s}
\gamma s \rangle_{\textrm{s.t.}}+\frac{1}{2}\langle \bar{u} \gamma_5
\gamma u \bar{s} \gamma_5 \gamma s \rangle_{\textrm{s.t.}},
\\
\overline{\mathcal{W}}_M^{\textrm{subt.}}[\Pi_{\Sigma^{+}(4q),u}(q_0^2,\vert\vec{q}\vert)]=&~\bar{\mathcal{B}}[\Pi^e_{{\Sigma^{+}(4q)},u}(q_0^2,\vert\vec{q}\vert)]_{\textrm{subt.}}-\bar{E}_{\Sigma^{+},q}\bar{\mathcal{B}}[\Pi^o_{{\Sigma^{+}(4q)},u}(q_0^2,\vert\vec{q}\vert)]_{\textrm{subt.}}\nonumber\\
=&~2\bar{E}_{\Sigma^{+},q}\Big[ \langle \bar{u} \gamma u  \bar{u}
\gamma u \rangle_{\textrm{s.t.}}-\langle \bar{u} \gamma_5 \gamma u
\bar{u} \gamma_5 \gamma u \rangle_{\textrm{s.t.}}+\langle \bar{u}
\gamma u \bar{s} \gamma s \rangle_{\textrm{s.t.}}-\langle \bar{u}
\gamma_5 \gamma u \bar{s} \gamma_5 \gamma s
\rangle_{\textrm{s.t.}}\Big].
\end{align}
The other OPE terms up to dimension 5 condensates can be found in
Ref.~\cite{Jin:1994bh}.

\section{Sum rule analysis}\label{sec4}

\begin{figure}
\includegraphics[width=8.2cm]{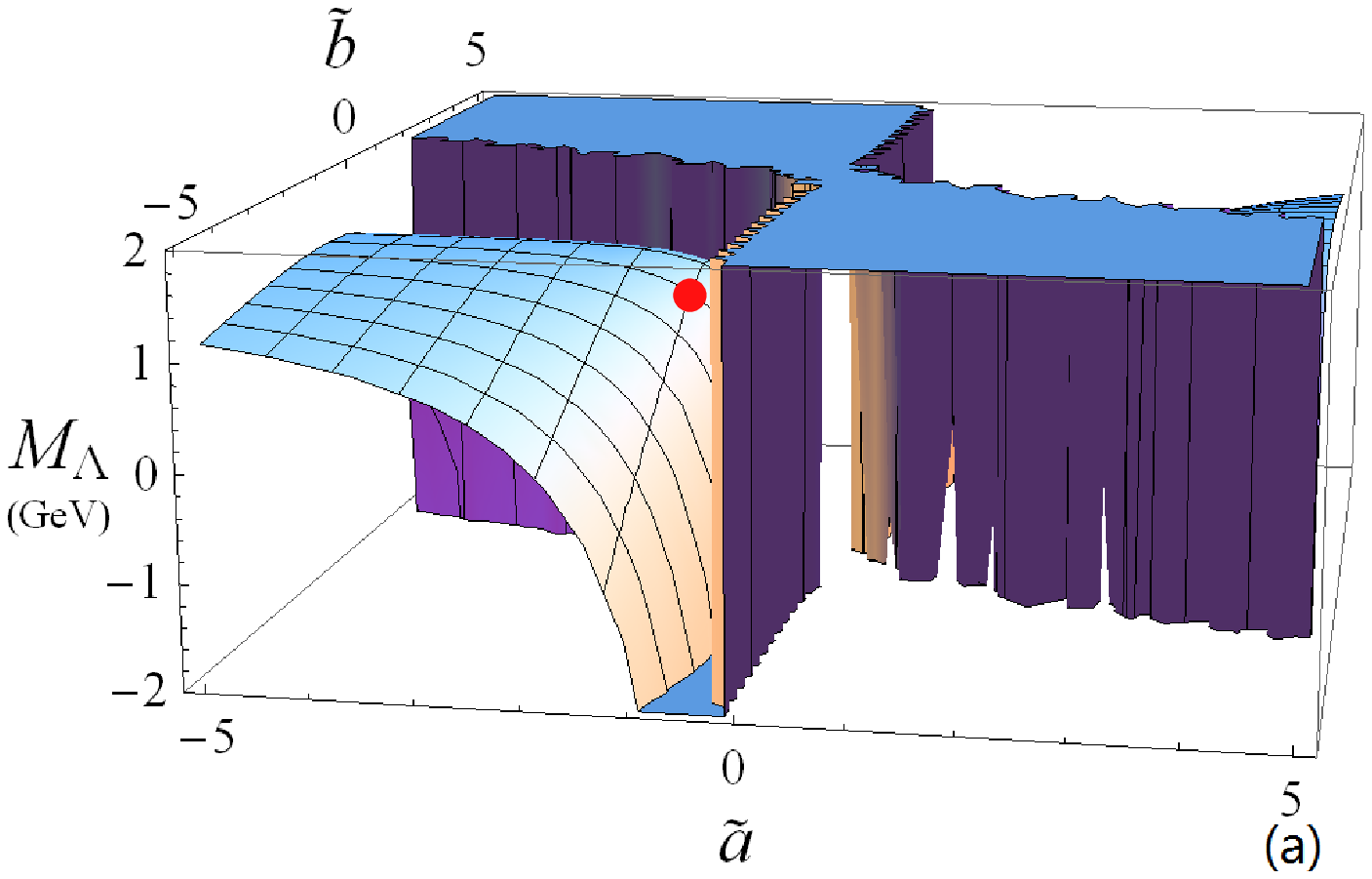}
\includegraphics[width=8.2cm]{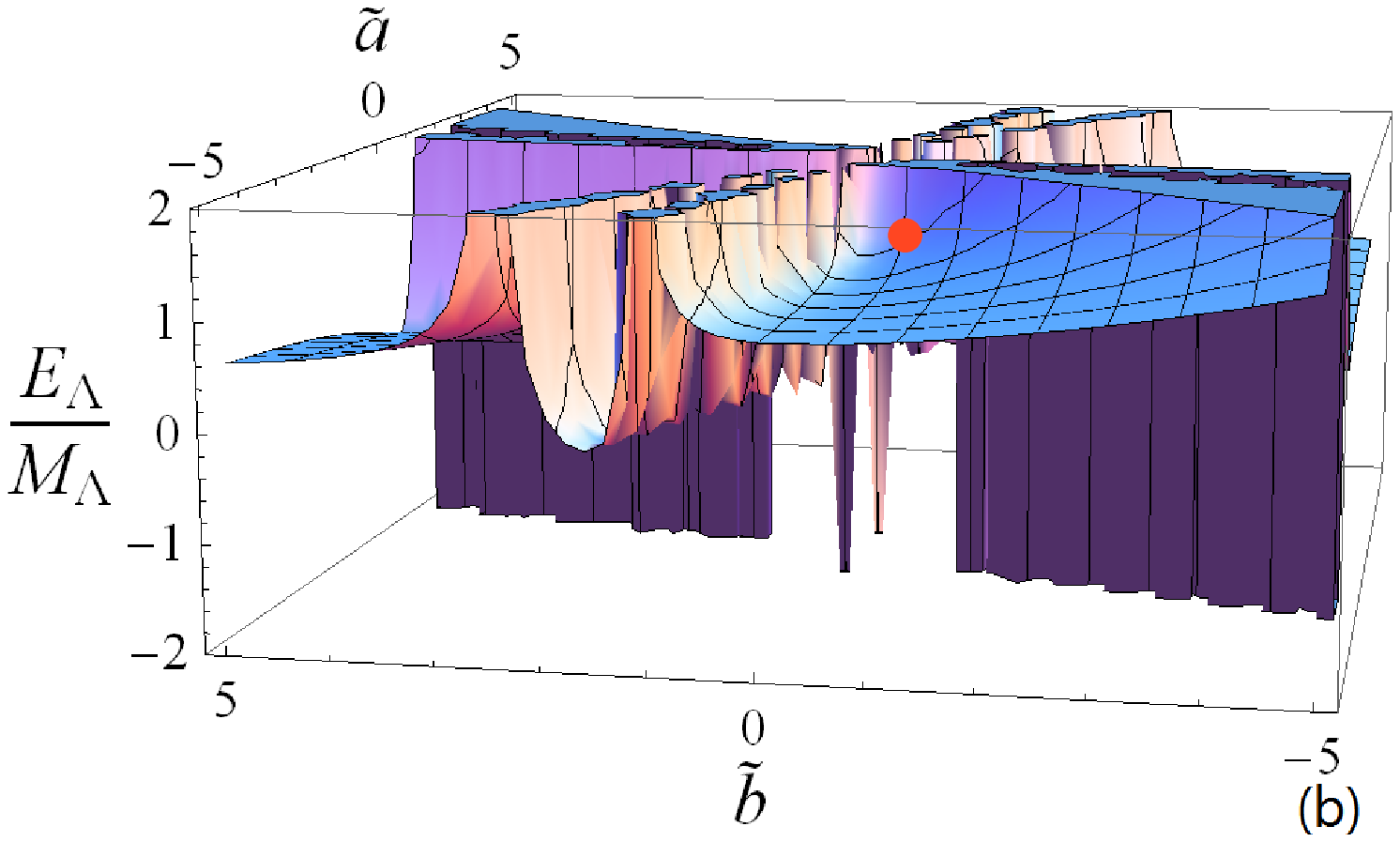}
\caption{Sum-rule result in  the $
\{\tilde{a},\tilde{b} \}$ plane for (a)  vacuum mass $M_\Lambda$,
and for (b) the ratio of the in-medium quasi-$\Lambda$ state energy
$E_\Lambda$ to $M_\Lambda$.    The point denoted by the red filled
circle corresponds to the result obtained with the Ioffe's choice
for the interpolating fields~\eqref{Ilcrnt}. Borel mass is set at
$M^2=1.1~\textrm{GeV}^2$.}\label{vcrnt}
\end{figure}

First of all, the two free parameters $\tilde{a}$ and $\tilde{b}$ in
the generalized $\Lambda$ interpolating fields \eqref{lcrnt} should
be confined before calculating the physical properties of the ground
resonance.  In Fig.~\ref{vcrnt}, one can find $\tilde{a}$ and
$\tilde{b}$ dependence of the vacuum mass $M_\Lambda$ and the ratio
between the medium energy of the quasi-$\Lambda$ state $E_\Lambda$
to $M_\Lambda$. In principle, if one can obtain the complete OPE of
the correlation function, the sum rules for the physical properties
should not depend on the choice of $\tilde{a}$ and $\tilde{b}$.
However, as we only have limited information for the condensates,
the OPE have to be truncated at some finite order and subsequently,
the sum rules can have singular and unstable region in the $\{\tilde{a},\tilde{b}\}$ plane. We assume that there is no
physically important singularity that appears only for a specific
linear combination of basis set \eqref{lcrnt} as the linear
combination corresponds to just a linear superposition of possible
spectral structures which commonly contain the same physical state.
Hence, small fluctuation in the coefficient space
$\{\tilde{a},\tilde{b}\}$ is possible and a reliable sum rules
should not drastically change by such a small fluctuation. As can be
found in Fig.~\ref{vcrnt}(a), the sum rule for vacuum mass becomes
unstable at $\vert \tilde{a} \vert \leq 1$ region so that the sum
rule with Ioffe's choice~\eqref{Ilcrnt} lies on the boundary of
stable region, which means that the calculated property can
drastically change by small variations in $\{\tilde{a},\tilde{b}\}$.
Within the same choice~\eqref{Ilcrnt}, the ratio $E_\Lambda
/M_\Lambda$ also lies on the unstable region and the value becomes
$\sim1.5$, which means that the quasi-$\Lambda$ state feels a strong
repulsive potential [Fig.~\ref{vcrnt}(b)]. The energy can be made
less repulsive by including the following additional derivative
expansion on spin-1 four-quark condensate:
\begin{align}
\langle \bar{s}\gamma_\mu s \bar{q}q\rangle_{\rho,I} &\Rightarrow
x^\nu\langle
\bar{s}\gamma_\mu D_\nu s \bar{q}q\rangle_{\rho,I},\\
\langle \bar{s}\gamma_\mu iD_\nu s \bar{q}q\rangle_{\rho,I} &=
\frac{1}{4}g_{\mu\nu}m_s\langle \bar{s}s \bar{q}q\rangle_{\rho,I}+
\frac{4}{3}\left(u_\mu u_\nu -
\frac{1}{4}g_{\mu\nu}\right)\left(\langle s^\dagger iD_0 s
\bar{q}q\rangle_{\rho,I}-\frac{1}{4}m_s\langle \bar{s}s
\bar{q}q\rangle_{\rho,I} \right),
\end{align}
where we used factorization \eqref{4qscalar2} for both the trace and
symmetric traceless parts. By adopting these additional steps and
condensates such as $\langle q^\dagger iD_0 q  \rangle_{\rho,I} $,
one can obtain a result similar to that of Ref.~\cite{Jin:1993fr}.
However, such derivative expansion and factorization may cause large
uncertainty and constitutes only part of the higher dimensional
contributions. Moreover, even if these artificial steps are
considered, with parameter set of $y\simeq 0.1$,  $f_1\ll1$, and
$f_2\simeq 1$, the quasi-$\Lambda$ state is still repulsive, not
consisted with the light-bounded state observed from the $\Lambda$
hyper-nuclei experiments \cite{Millener:1988hp}.

\begin{figure}
\includegraphics[width=6.4cm]{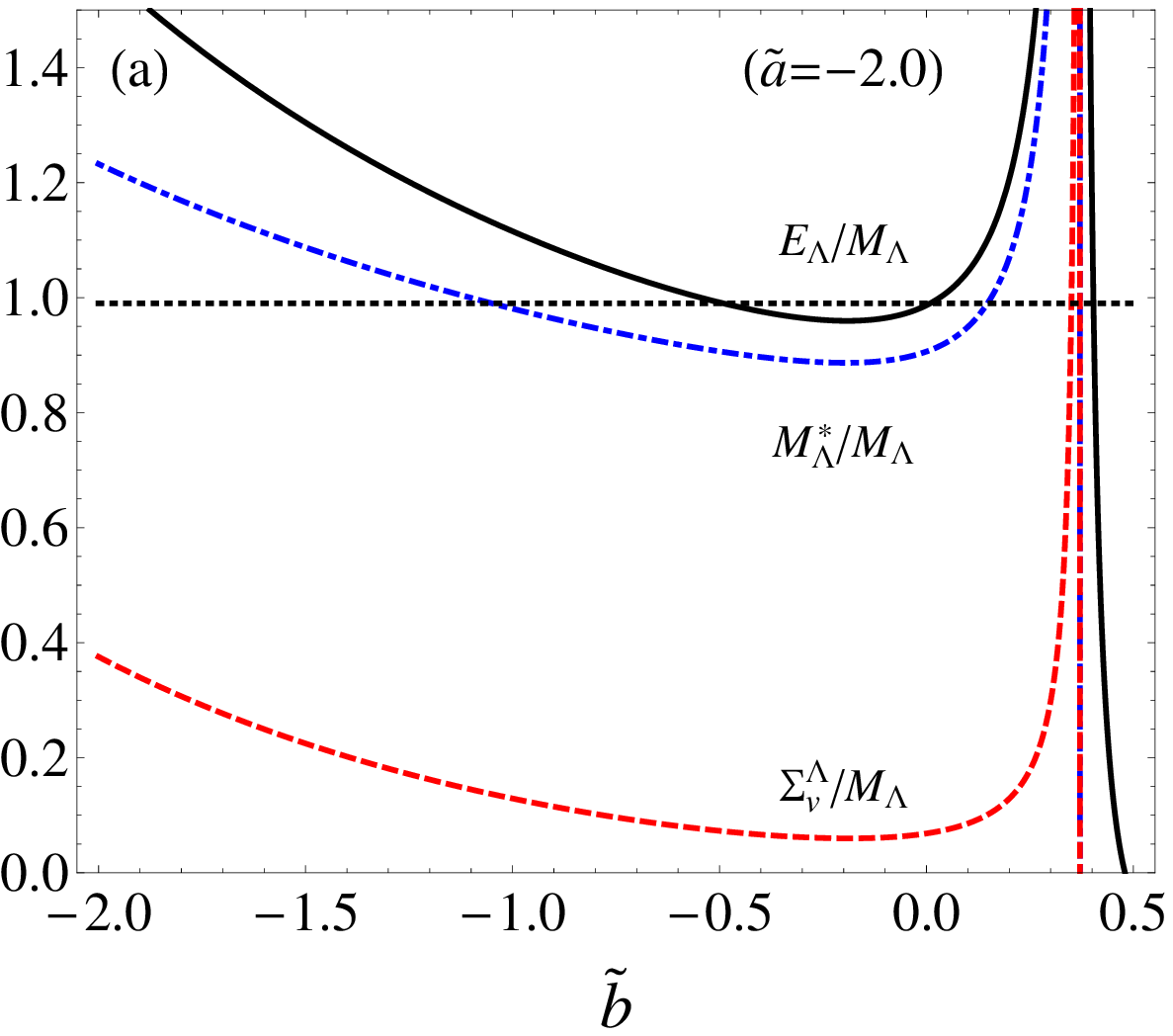}
\includegraphics[width=6.4cm]{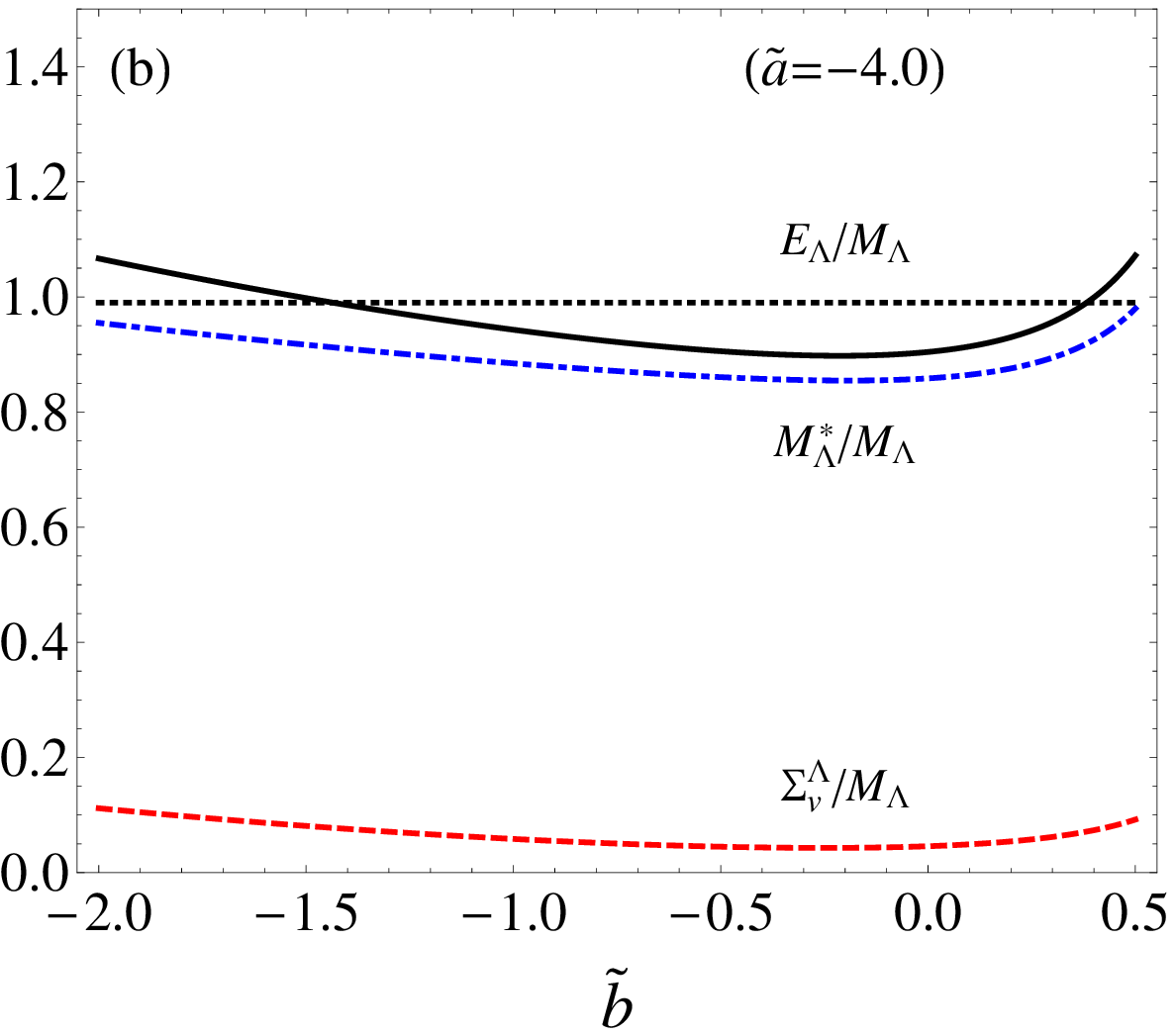}
\caption{Ratios of the in-medium quasi-$\Lambda$ state energy, scalar, and vector self energies to its vacuum mass for $\tilde{a}=$ -2.0 (left) and -4.0 (right) as a function of $\tilde{b}$.  Black dotted line represents
0.99.
}\label{crosssec}
\end{figure}

Therefore $\tilde{a}$ and $\tilde{b}$ should be redefined to ensure
small variation of sum rules in the ``small perturbation'' in
$\{\tilde{a},\tilde{b}\}$ plane. As one can find in
Fig.~\ref{vcrnt}(b), the quasi-$\Lambda$ pole becomes stable and
bounded in large $ \vert \tilde{a} \vert $. This tendency can be
clearly found from the cross-sections in Fig.~\ref{vcrnt} for a
fixed $\tilde{a}$  given in Fig.~\ref{crosssec}.
The stable range of $\tilde{b}$ for the quasi-$\Lambda$ pole appears
from $\tilde{a}\sim -2$ but the range is still narrow
[Fig.~\ref{crosssec}(a)]. In the region where $ \vert \tilde{a}
\vert $ is large [Fig.~\ref{crosssec}(b)], one finds that the
stable range of $\tilde{b}$ becomes wider. The stable range of
$\tilde{b}$ can be identified as $-0.5 < \tilde{b} < 0$ at $ \vert
\tilde{a} \vert \gg 2 $. We sampled the 9 stable points in
$\{\tilde{a},\tilde{b}\}$ plane and averaged the sum rules from
these points:
\begin{align}
\{(\tilde{a},\tilde{b})\}=
\big\{&(-1.80,-0.10),(-1.80,-0.15),(-1.80,-0.20),\nonumber\\
&(-2.00,-0.05),(-2.00,-0.15),(-2.00,-0.25),\nonumber\\
&(-2.20,-0.00),(-2.20,-0.15),(-2.20,-0.30)\big\},\label{sppt}
\end{align}
where the central point is $\{-2.00,-0.15\}$. If one takes the factorization constant $k_1$ less than 1,  the central point of the stable region should be located at $\tilde{a}<-2$.

\begin{figure}
\includegraphics[height=6.5cm]{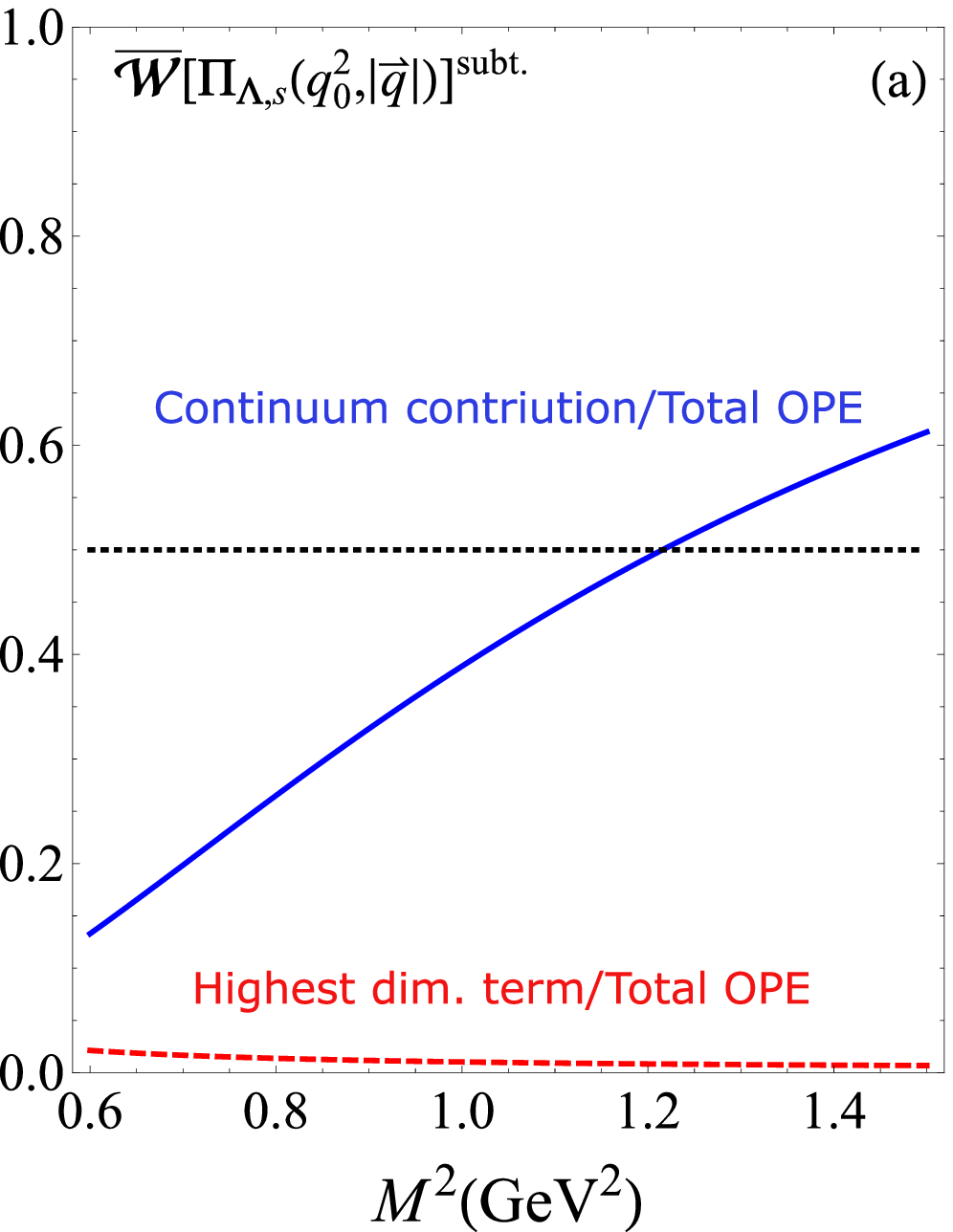}
\includegraphics[height=6.5cm]{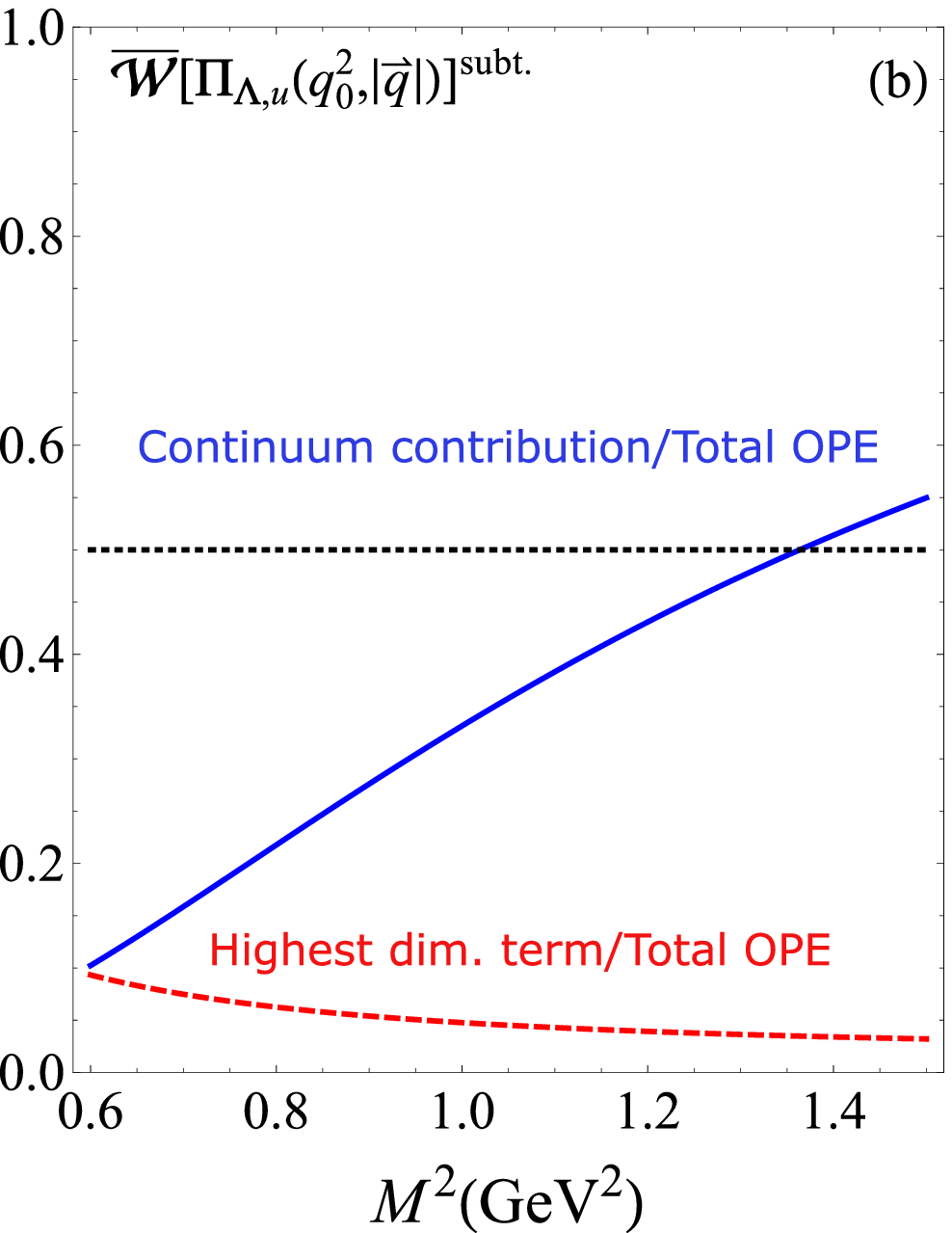}
\includegraphics[height=6.5cm]{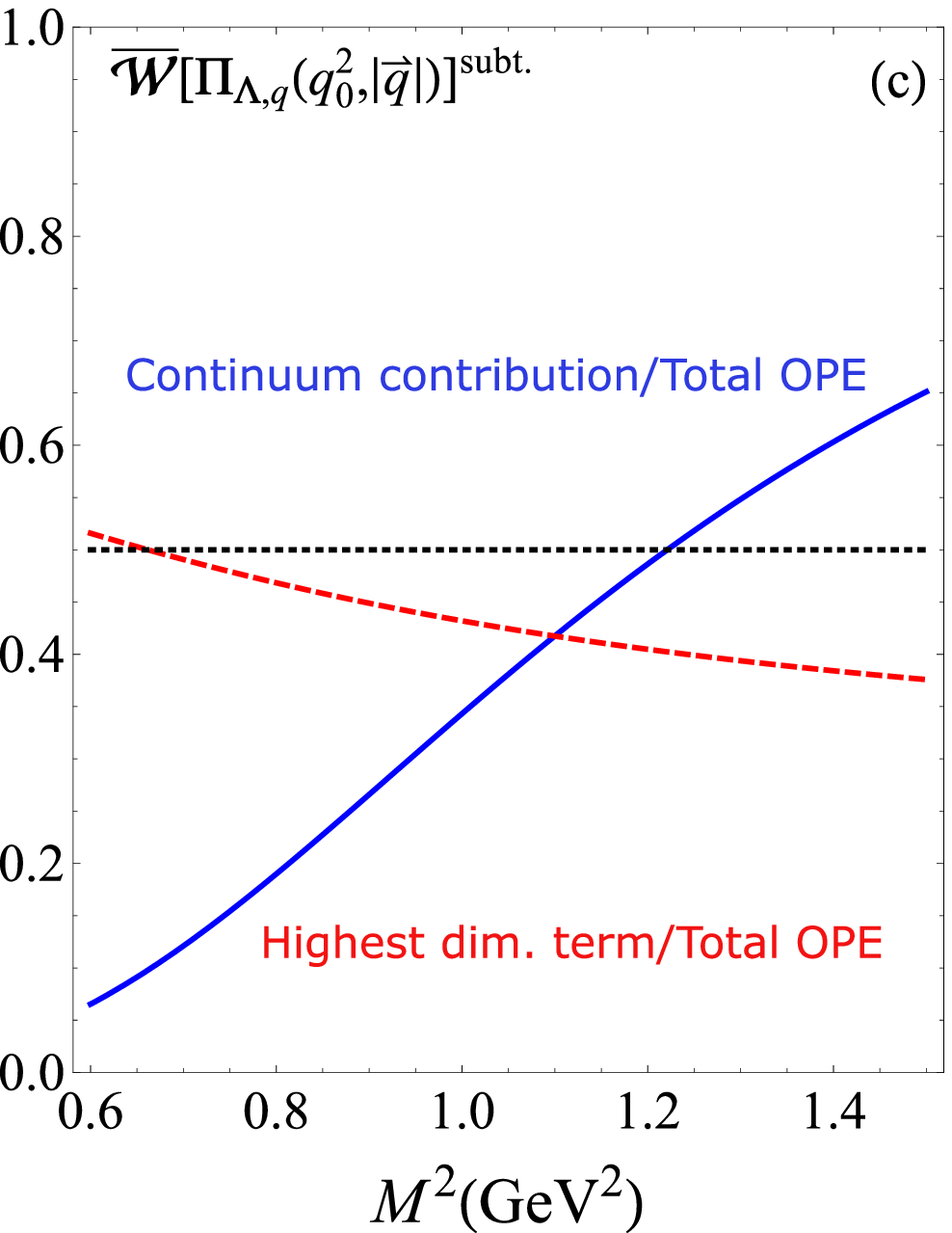}
\caption{Ratios of the Borel-transformed-continuum contribution and highest-dimensional condensate terms to the Borel transformed subtracted total OPE of the $\Lambda$ for the three different invariants (a) $\Pi_{\Lambda,s}$, (b) $\Pi_{\Lambda,u}$, and (c) $\Pi_{\Lambda,q}$.
The black dotted line represents 50\%.}\label{BW}
\end{figure}

The next task is identifying a reliable range of the Borel mass.
Again, if one could obtain ``the complete OPE'' of each
invariant, the physical observable should not depend on Borel mass.
However, as the OPE is truncated, the reliable sum rules should be
found through the specific range of Borel mass called the Borel
window. We used the following simple criteria: (i) the continuum
contribution should not exceed 50\% of the total OPE contribution
and (ii) the highest mass dimensional condensate terms should not
exceed 50\% of the total OPE contribution. If the OPE of each
invariant is a well constructed asymptotic series, the sum rules
should show ``plateau'' or at least very weak dependence in the
Borel mass. In Fig.~\ref{BW}, one can find that the Borel window can
be set as $1.0~\textrm{GeV}^2 \leq M^2 \leq1.2~\textrm{GeV}^2$. As
can be found in the latter part, the sum rules are almost independent
on Borel mass in this window.

\begin{figure}
\includegraphics[height=6.5cm]{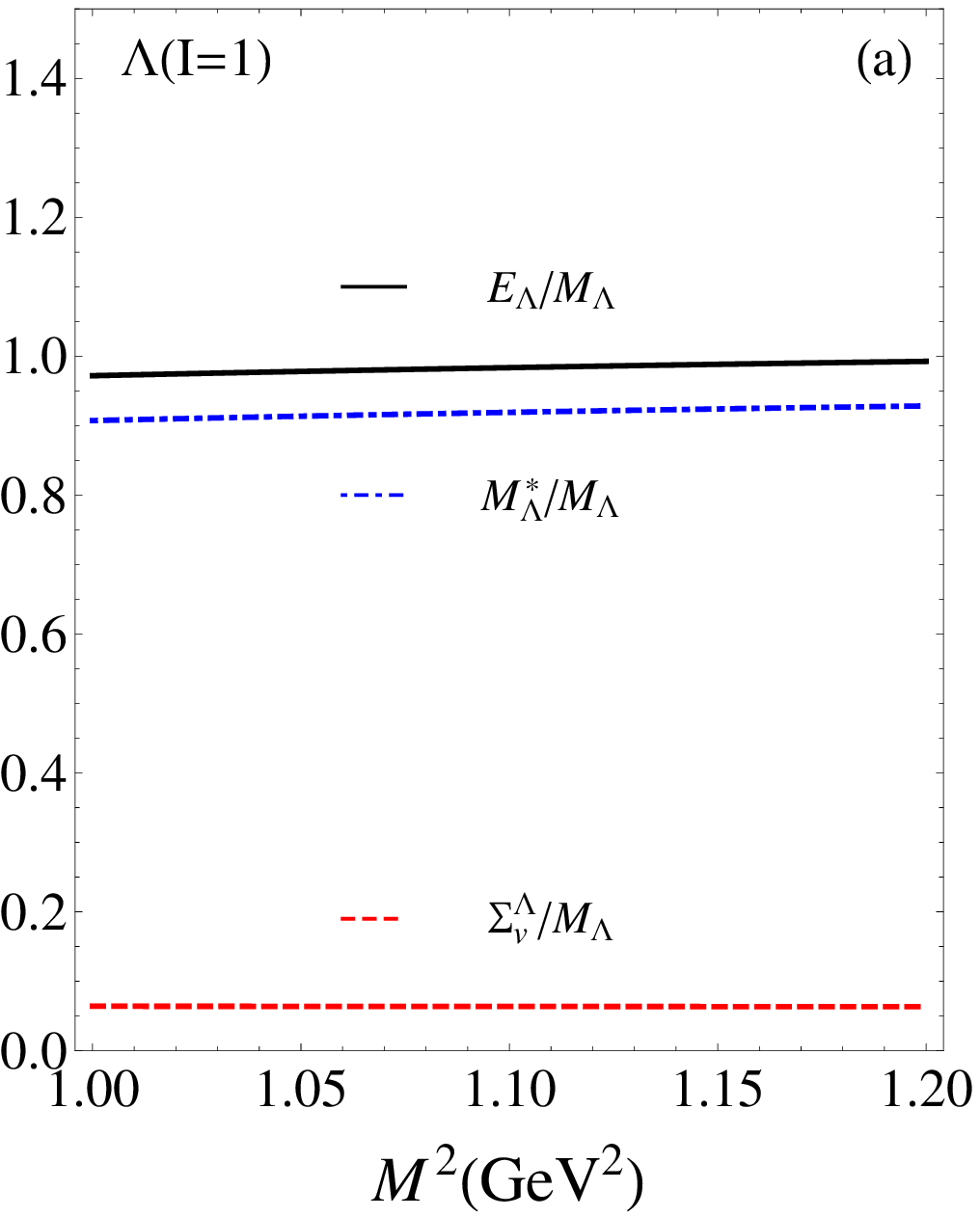}
\includegraphics[height=6.5cm]{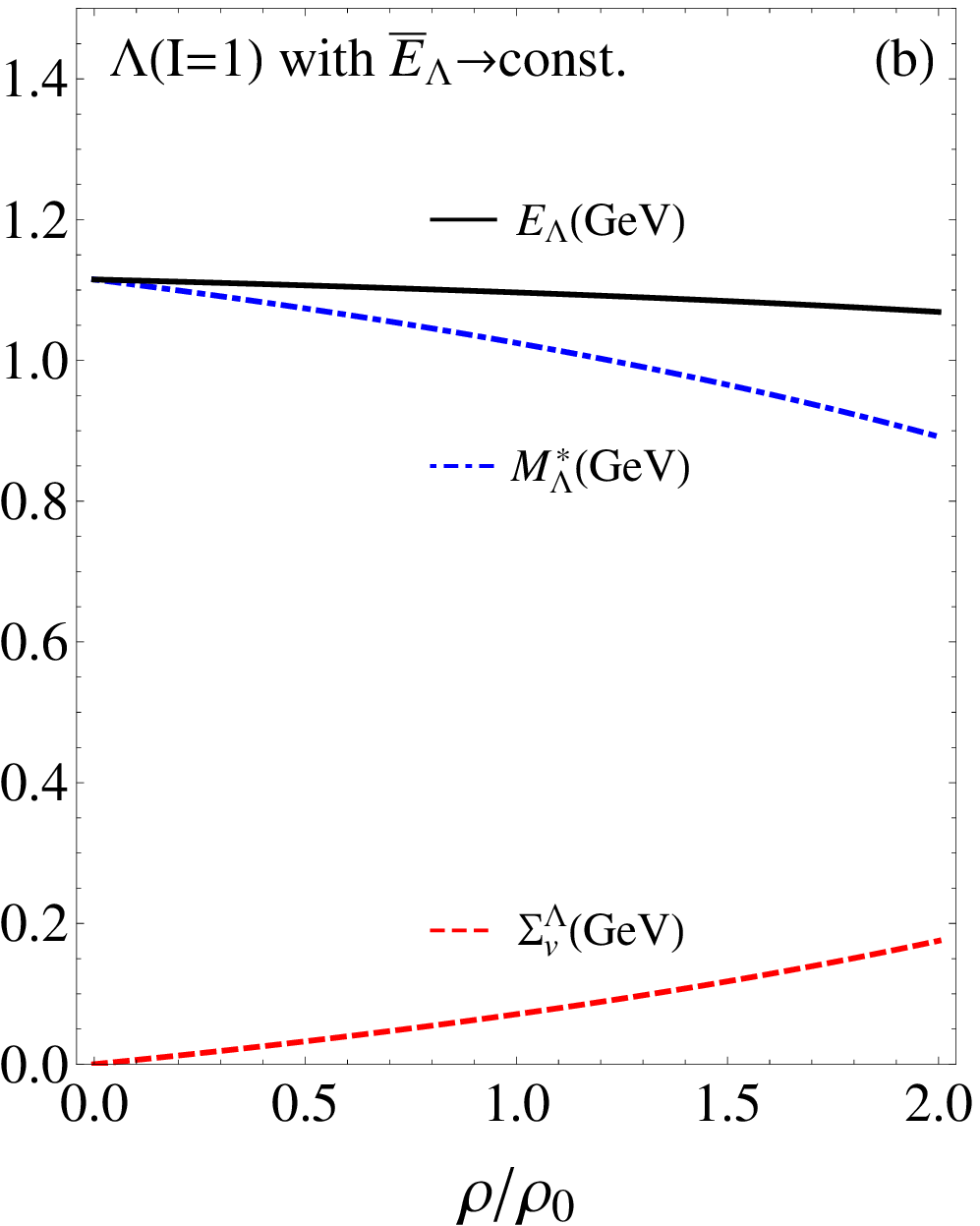}
\includegraphics[height=6.5cm]{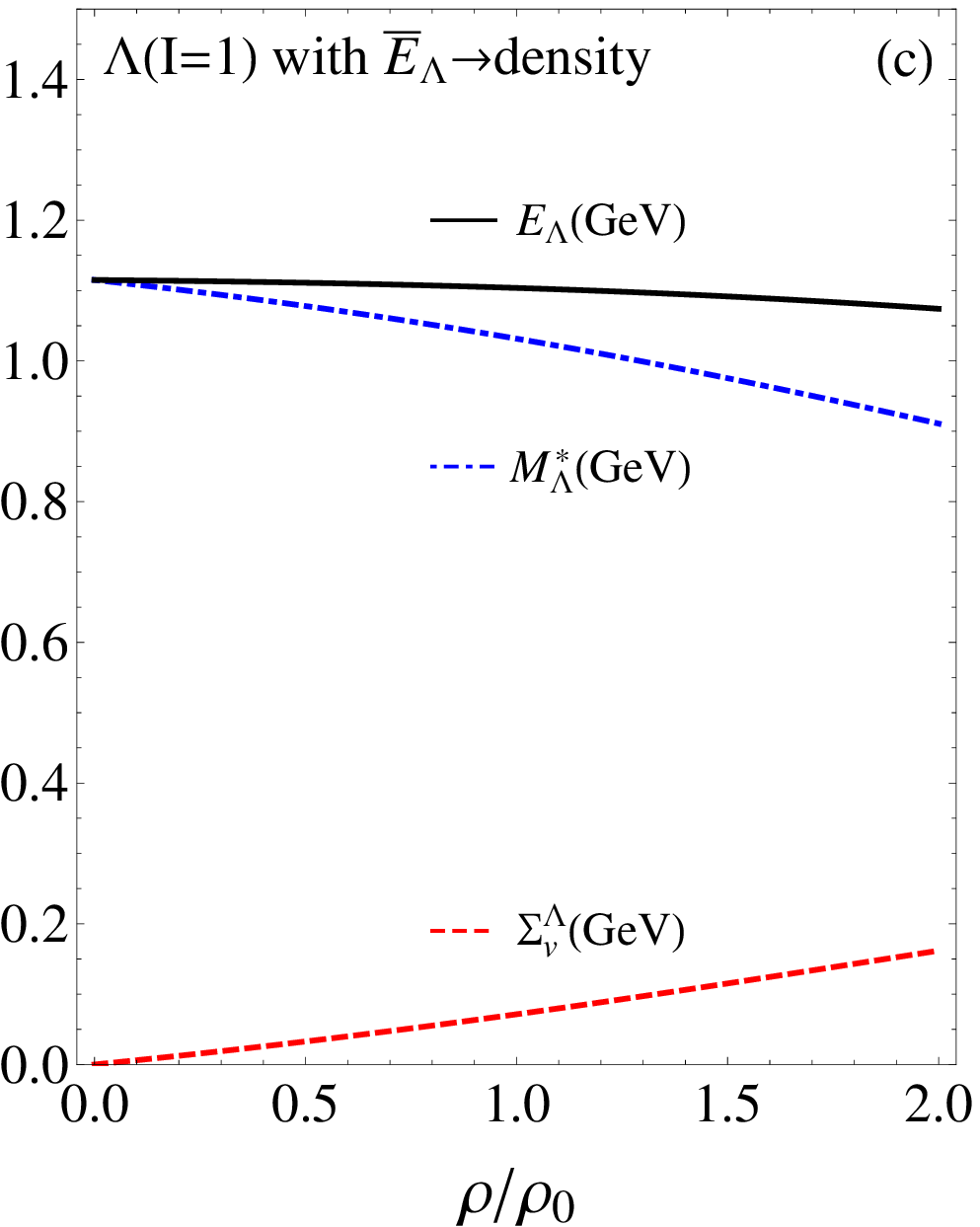}
\caption{(a) Ratios and (b), (c) sum rules for the quasi-$\Lambda$
self-energies and the quasi-$\Lambda$ pole. $\rho=\rho_0$ in graph
(a). In the linear density approximation, in-medium $\Lambda$ sum rules
do not depend on iso-spin asymmetry of the surrounding matter. Graph
(b) has been plotted with constant quasi-anti $\Lambda$ pole and
graph (c) has been plotted with density dependent quasi-anti
$\Lambda$ pole. Units for the graphs in (b) and (c) are GeV.}\label{lsr}
\end{figure}

The sum-rule results for the quasi-$\Lambda$ state have been plotted
in Fig.~\ref{lsr}. The scalar attraction has been found as
$M^{*}_\Lambda/M_\Lambda \simeq 85\%$ and the vector repulsion as
$\Sigma^\Lambda_v/M_\Lambda \leq 10\%$ [Fig.~\ref{lsr}(a)].
Comparing with the results \cite{Jin:1993fr} where Ioffe's choice
has been used, our results show much weaker strength of attraction
and repulsion. Comparing the strength of the self-energies with the
self-energies of the nucleon sum rules \cite{Drukarev:1988kd,
Cohen:1991js, Furnstahl:1992pi, Jin:1992id, Jin:1993up,
Cohen:1994wm, Jeong:2012pa}, one can find the ratio as
$\Sigma_{s,\Lambda}/\Sigma_{s,N}\simeq 0.31$ and
$\Sigma_{v,\Lambda}/\Sigma_{v,N}\simeq 0.26$,  almost $30\%$ of the
nucleon case. It means that the naive valence quark number counting
may not be good for the determination of interaction strength
between nucleon and  hyperon. The net effect, estimated from the
ratio $E_\Lambda /M_\Lambda$, is an attraction in the nuclear matter
within 10 MeV scale. As the spin-orbit coupling in $\Lambda$
hyper-nuclei is expected to be weak \cite{Brockmann:1977es,
Hiyama:2000jd}, the mean-field type phenomenology should explain the
experimental observation of weakly bounded quasistate, which is now
well reproduced in the sum-rule approach.

\begin{figure}
\includegraphics[height=6.5cm]{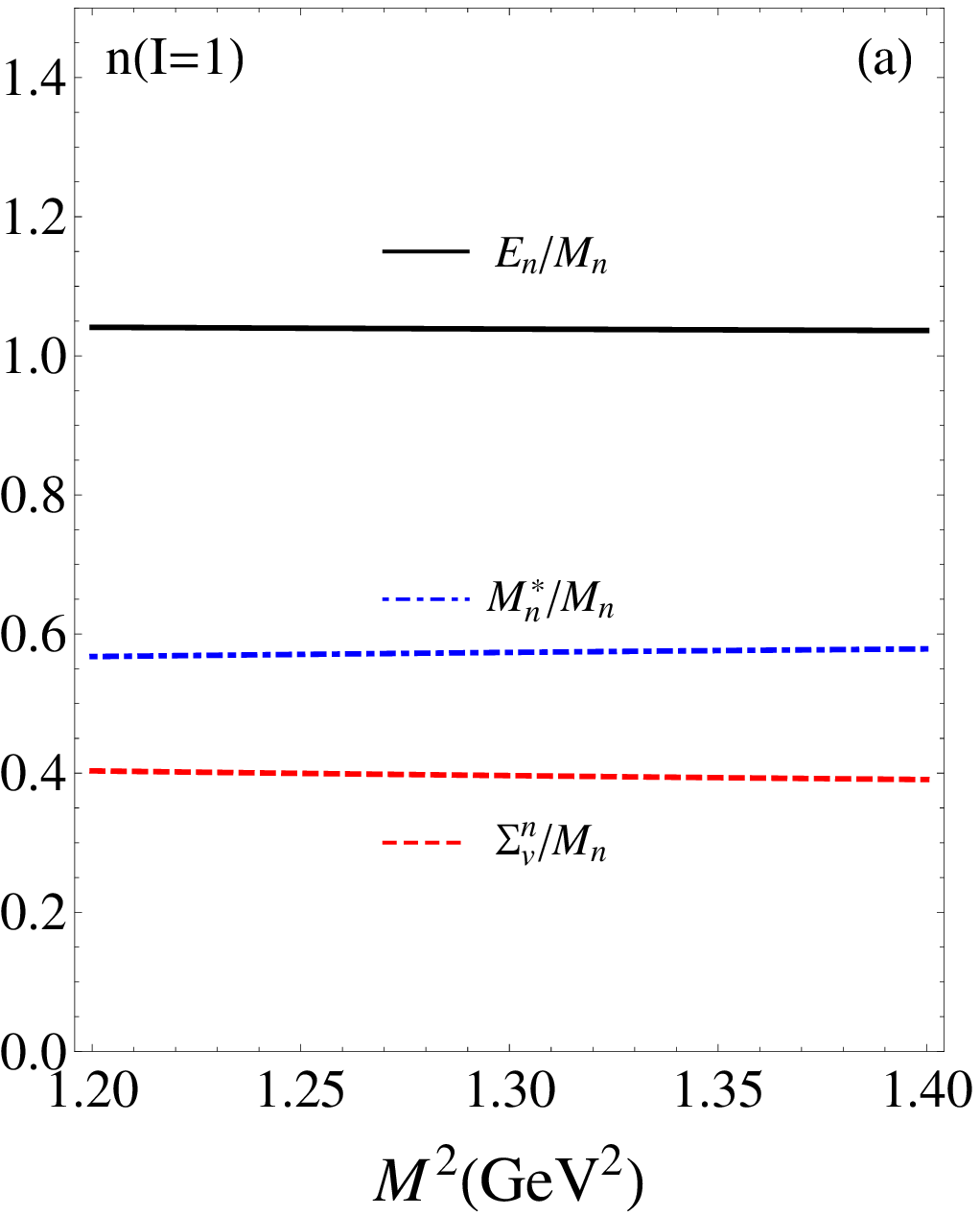}
\includegraphics[height=6.5cm]{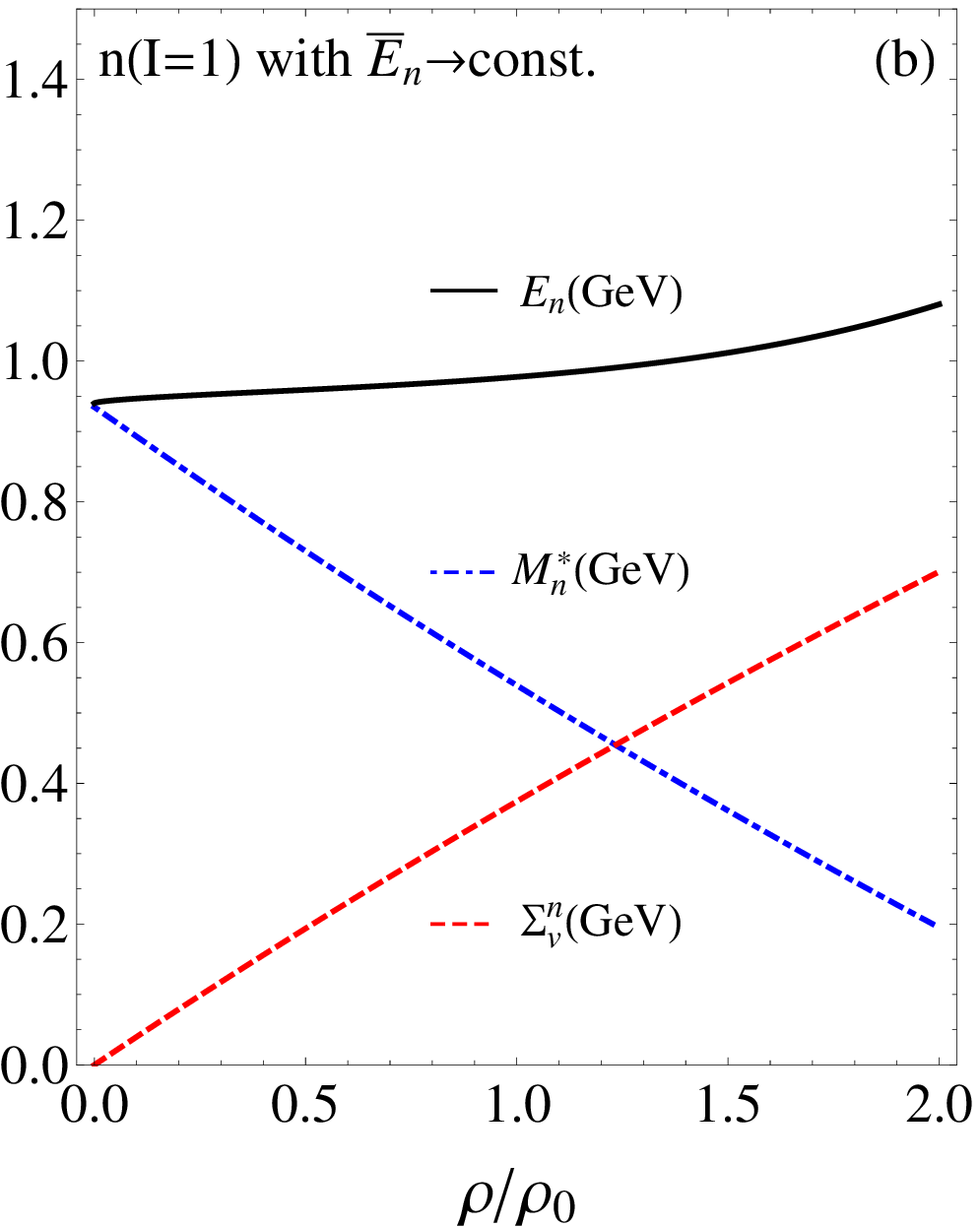}
\includegraphics[height=6.5cm]{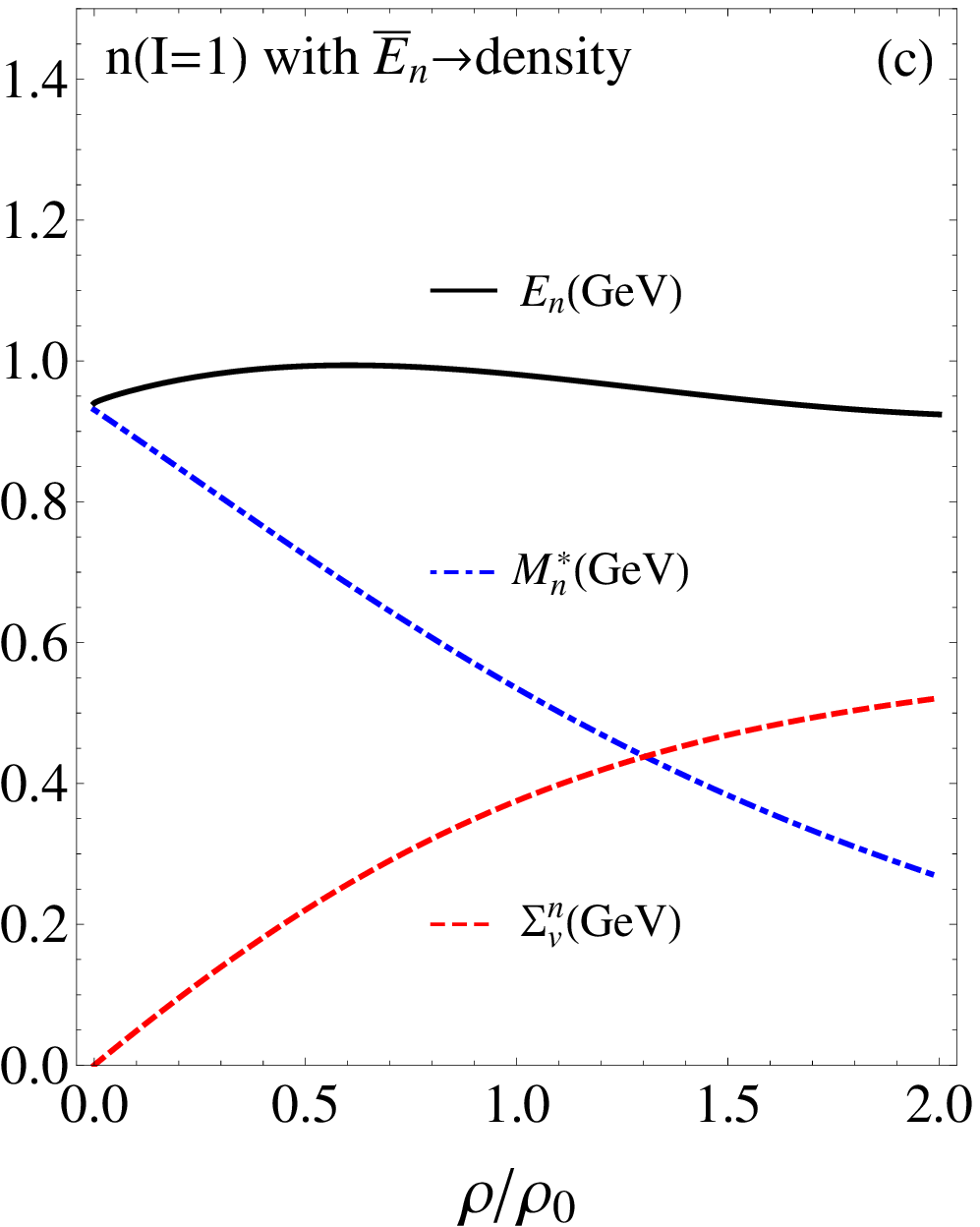}
\caption{(a) Ratios and (b), (c) sum rules for neutron self-energies and the
quasineutron pole in the neutron matter [$\rho=\rho_0$ in graph
(a)]. Graph (b) has been plotted with constant quasi-antineutron
pole and graph (c) has been plotted with density dependent
quasi-antineutron pole. Units for the graphs in (b) and (c) are
GeV.}\label{nsr}
\end{figure}

\begin{figure}
\includegraphics[height=6.5cm]{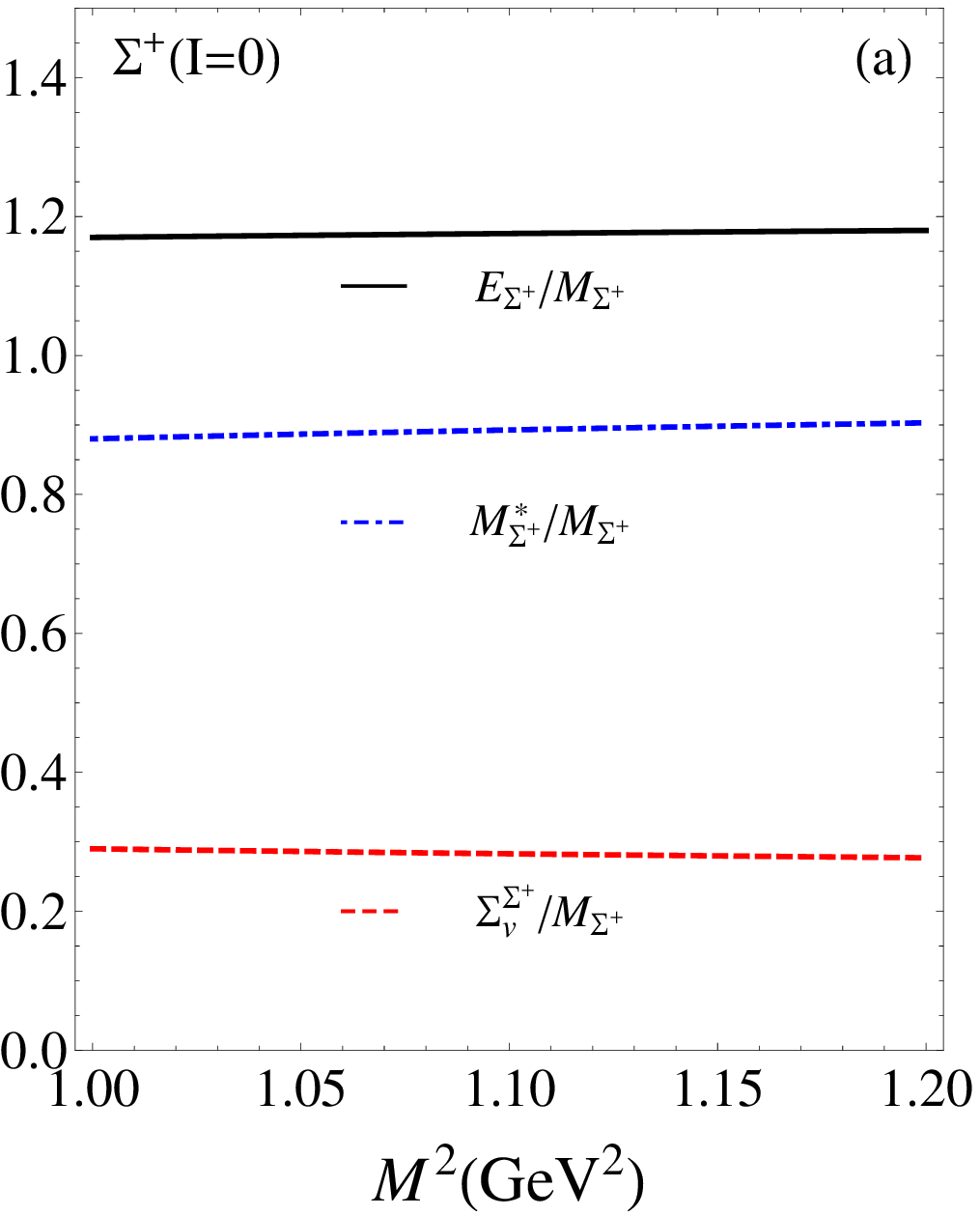}
\includegraphics[height=6.5cm]{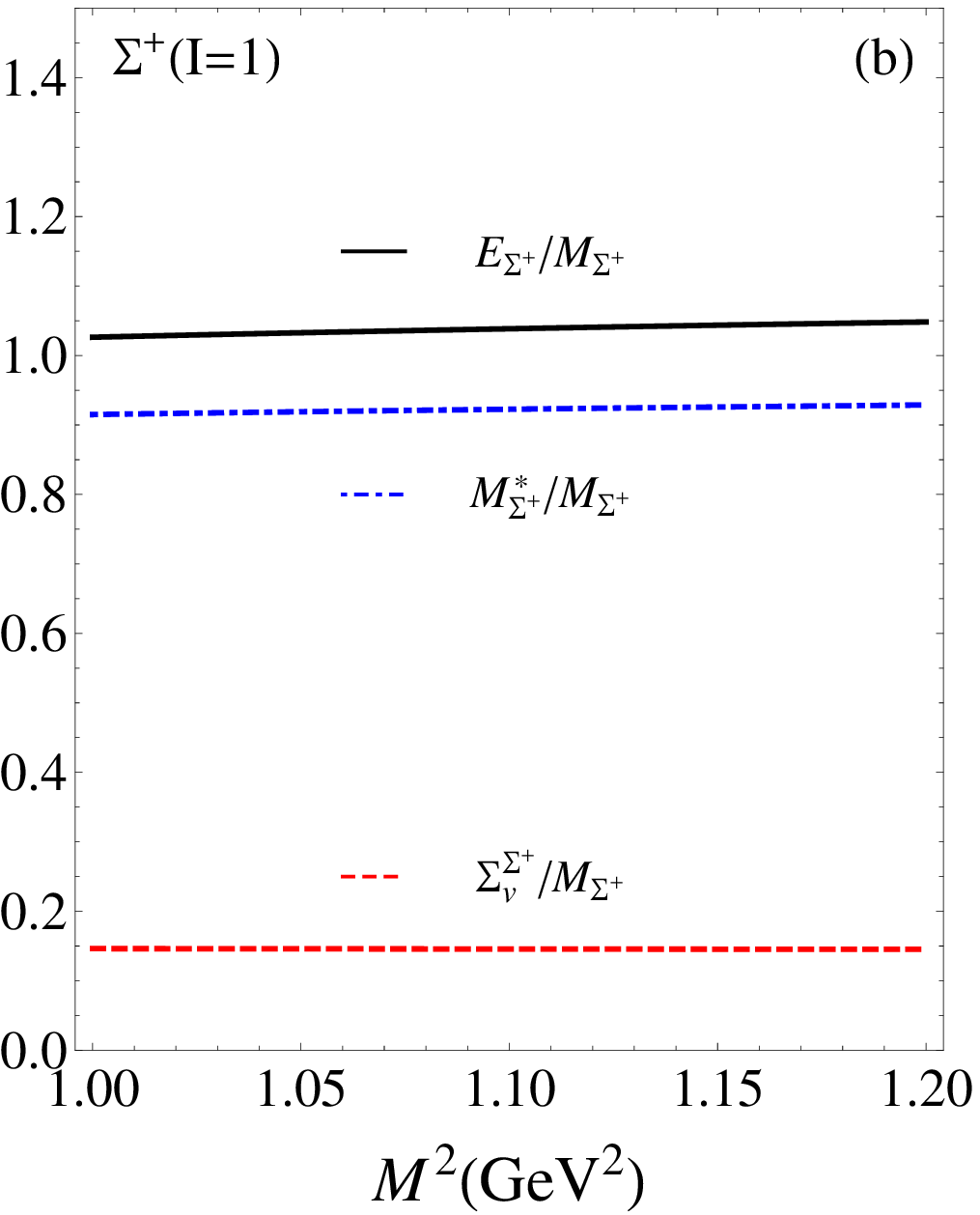}
\includegraphics[height=6.5cm]{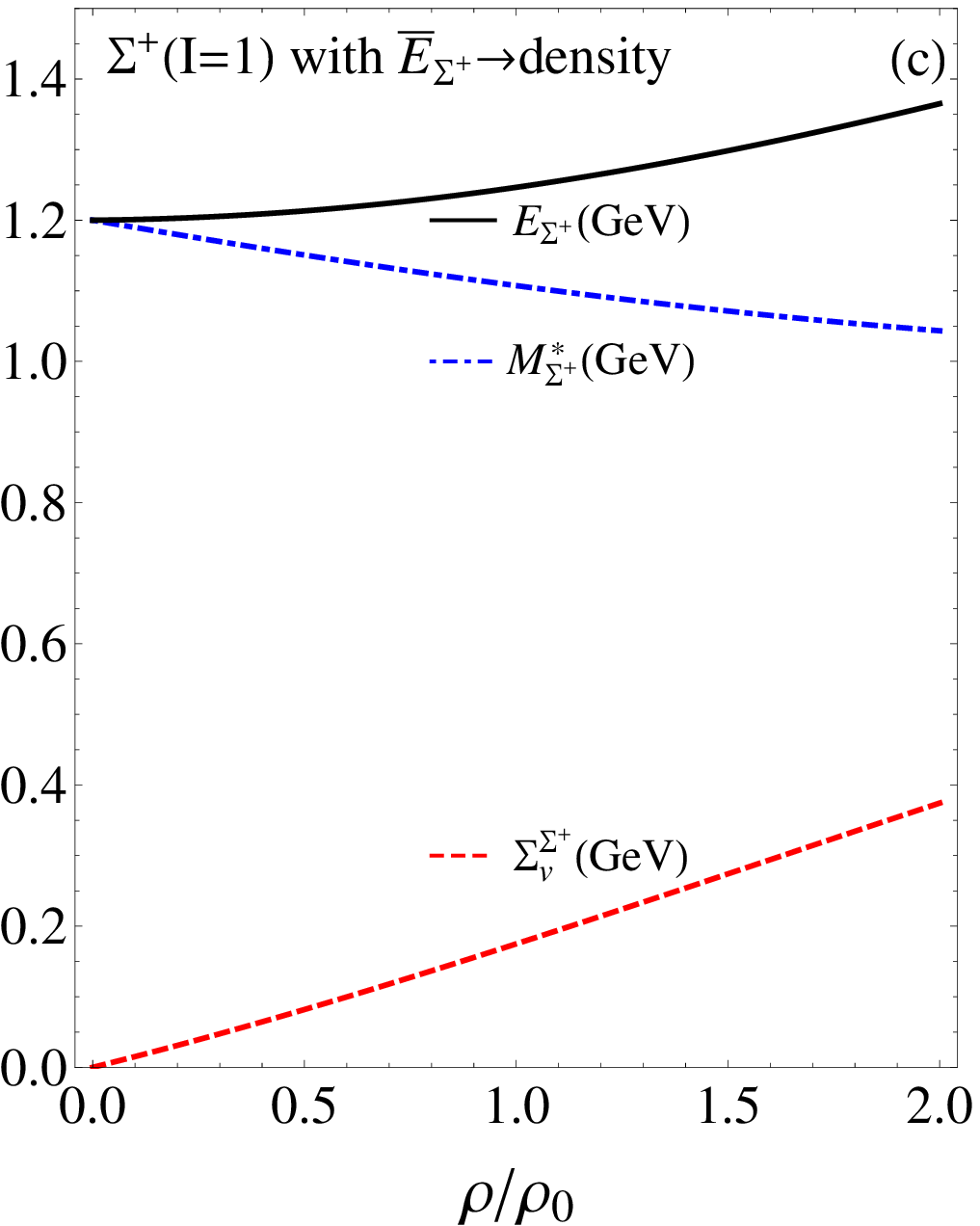}
\caption{(a), (b) Ratios and (c) sum rules for $\Sigma^{+}$ self-energies and
the quasi-$\Sigma^{+}$ pole. Graph (a) is plotted in iso-spin-symmetric condition ($I=0$, $\rho=\rho_0$) and graph (b) is plotted
in neutron matter condition ($I=1$, $\rho=\rho_0$). For the
in-medium $\Sigma^{+}$ sum rules, it is negligible for the
difference between the constant and density-dependent quasi-anti
$\Sigma^{+}$ pole case. Units for the graph in panel (c) are GeV.}\label{ssr}
\end{figure}

As discussed in detail in Appendix~\ref{appentw}, we will consider
the most general form for the twist-4 matrix elements in this work
allowing for matrix elements not considered previously in
Ref.~\cite{Jeong:2012pa}. Hence, the nucleon sum rules should be
reexamined. The nucleon sum rules in the iso-spin-symmetric
condition ($I=0$) do not show significant difference from the
results of Ref.~\cite{Jeong:2012pa}. For the neutron matter ($I=1$)
case, each sum rule for neutron and $\Sigma^{+}$ with renewed
four-quark OPE and twist-4 matrix elements is plotted in
Figs.~\ref{nsr} and~\ref{ssr}, respectively. The quasi-neutron
pole is slightly repulsive with the scalar attraction $M^{*}_n/M_n
\simeq 55\%$ and the vector repulsion $\Sigma^n_v/M_n \simeq 40\%$
[Fig.~\ref{nsr}(a)]. If one regards the quasi-antipole as a given
constant, the quasineutron pole monotonically increases but if the
pole has the density dependence, it decreases after
$\rho/\rho_0\sim0.6$ [Figs.~\ref{nsr}(b) and~\ref{nsr}(c)].
This density behavior is only reliable near $\rho/\rho_0\sim1$;
$0.5<\rho/\rho_0<1.5$.

Among the $\Sigma$ family, we have calculated the in-medium
$\Sigma^{+}$ sum rules as it is expected to have lowest
quasiparticle energy in the neutron matter when the electromagnetic interaction is neglected. As one can find in
Fig.~\ref{ssr}, sum rules show a weak scalar attraction
$M^{*}_{\Sigma^{+}}/M_{\Sigma^{+}} \simeq 90\%$, a strong vector
repulsion $\Sigma^{\Sigma^{+}}_v/M_{\Sigma^{+}} \simeq 30\%$ and a
strong net repulsion $E_{\Sigma^{+}} /M_{\Sigma^{+}} \simeq 120\%$
in the iso-spin-symmetric ($I=0$) condition. Then the total
repulsion exists in the order of 100 MeV scale, which fits with the
experimental observation from $\Sigma$ hyper-nuclei \cite{Noumi:2001tx}. In the
neutron matter condition ($I=1$), the vector repulsion becomes
weaker $\Sigma^{\Sigma^{+}}_v/M_{\Sigma^{+}} \simeq 15\%$ and the
net repulsive effect reduces $E_{\Sigma^{+}} /M_{\Sigma^{+}} \simeq
105\%$. Although the repulsion becomes weaker in the neutron matter,
the quasi-$\Sigma^{+}$ energy monotonically increases
[Fig.~\ref{ssr}(c)] and it never crosses with the quasineutron
energy at least in the region $0.5 <\rho/\rho_0<1.5$.

\begin{figure}
\includegraphics[height=6.5cm]{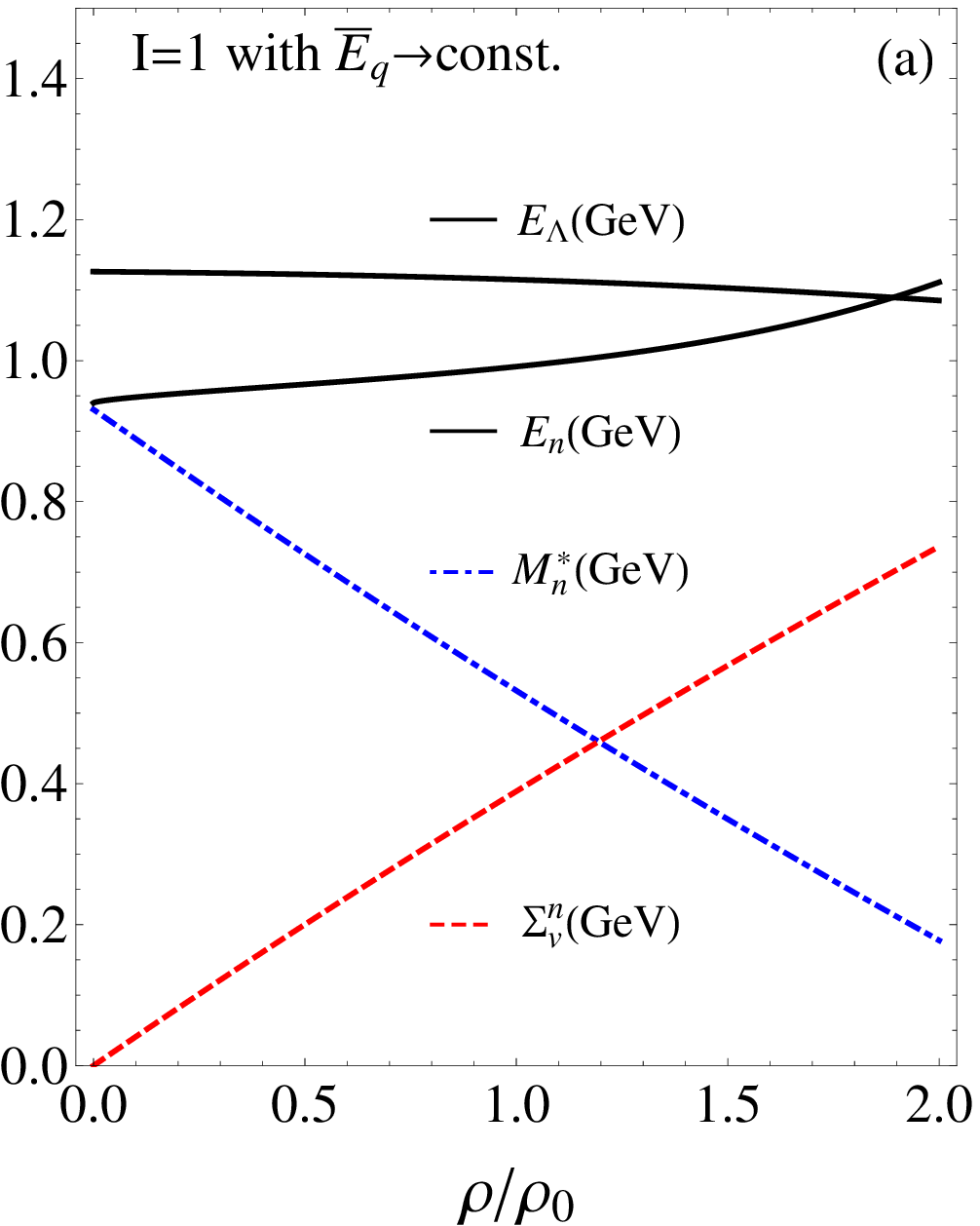}
\includegraphics[height=6.5cm]{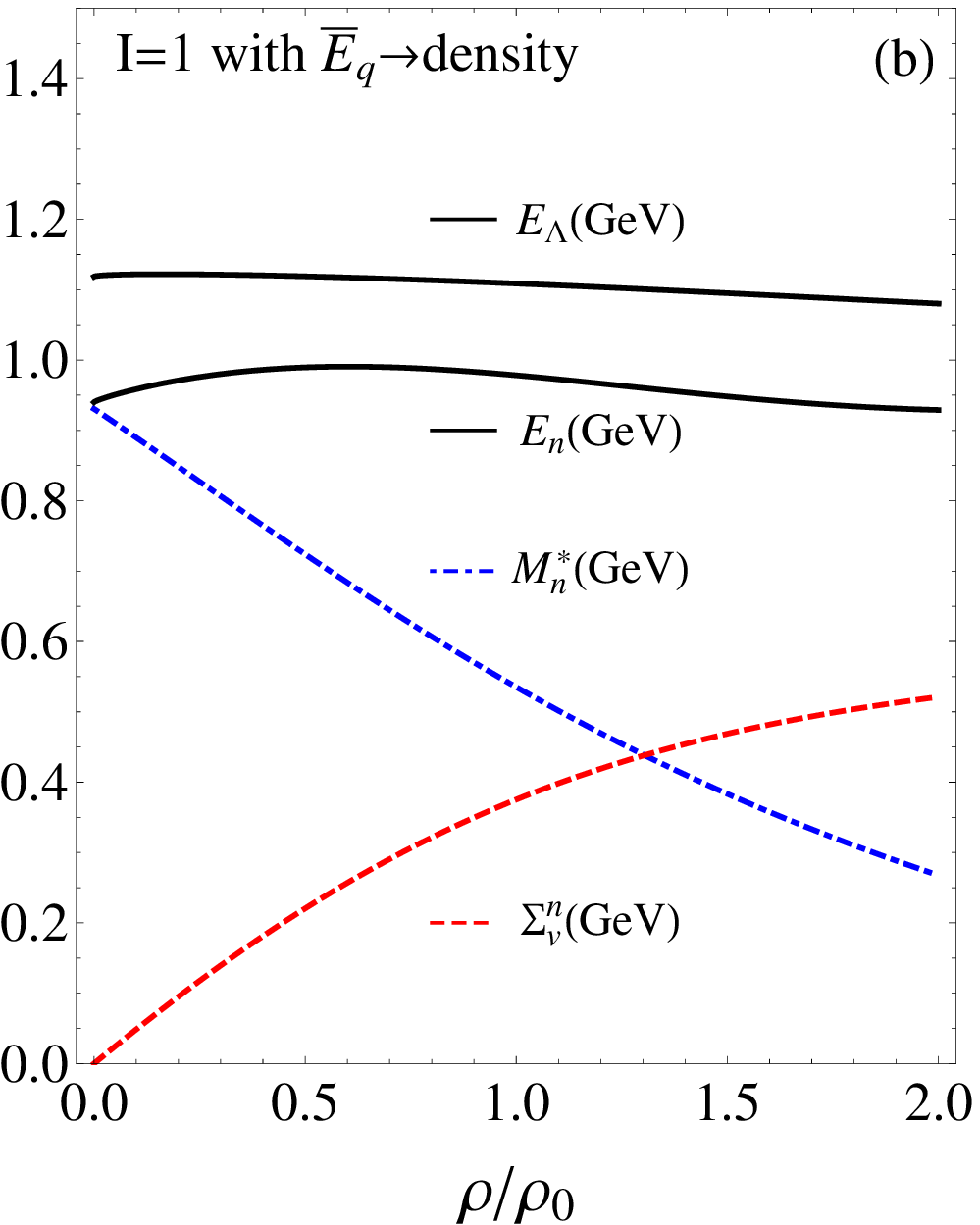}
\includegraphics[height=6.5cm]{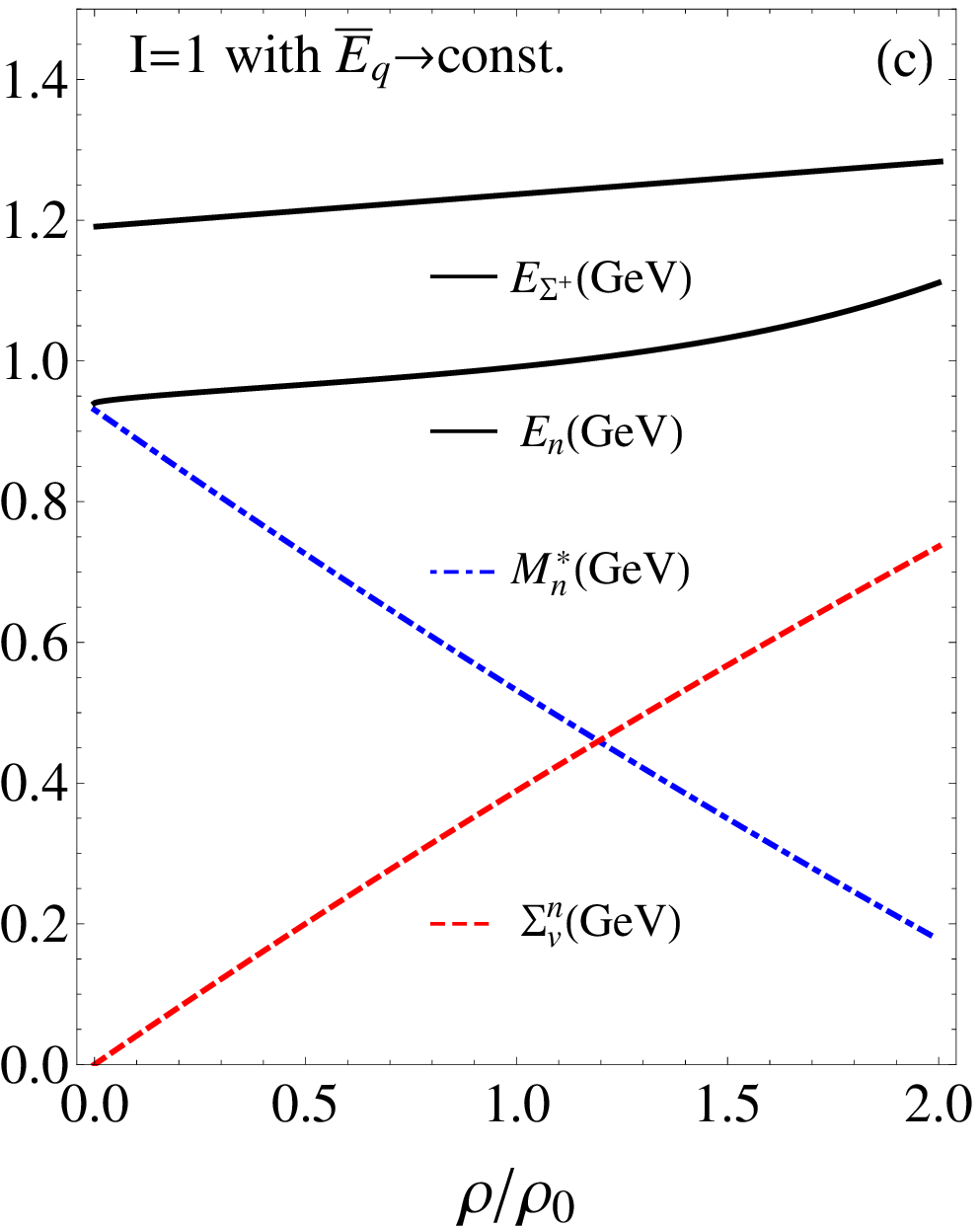}
\caption{Comparison of the density behavior between
$\Lambda$, $\Sigma^{+}$ hyperon and neutron sum rules in the neutron
matter. Only graph (a) where constant quasi-antipoles are assigned
shows the cross of quasibaryon energy. Units are GeV.}\label{nld}
\end{figure}

To discuss whether hyperons appear in the nuclear matter,  the
density behavior of the quasibaryon states should be compared with
each other. The density behavior is plotted in Fig.~\ref{nld}. In
the plots, only the quasi-$\Lambda$ state has a possibility to be
lower in energy than the quasineutron state in the neutron matter.
In the case where the quasi-antipole is just a given constant, the
quasi-$\Lambda$ pole crosses with the quasineutron pole at $\rho/
\rho_0\simeq1.8$. In the other cases, including the density behavior
of $\Sigma^{+}$, the crossing never occurs: the $\Sigma$ family
($I=1$) could be excluded in the discussion for early appearance of
hyper-nuclear matter problem. Because our sum rules are only reliable
near the normal density region $0.5 <\rho/\rho_0<1.5$, we can
conclude that the early crossing of the quasineutron and hyperon do
not occur in the reliable region so that the onset of hyper-nuclear
matter at low density becomes unlikely. To extend our result to the higher-density region, the density dependence for the condensates
should be known beyond the linear density approximation.

\section{Discussion and Conclusions}\label{sec5}

Starting from the most general interpolating fields without
derivatives for the nucleon, $\Sigma$ and $\Lambda$ hyperon as given
in Sec.~\ref{sec2}, we obtained the optimal set in the parameter
space for the interpolating fields that gives the most stable sum
rules. The famous construction scheme known as Ioffe's choice for
interpolating fields would be suitable for the nucleon and the
$\Sigma$ hyperon. However, for the $\Lambda$ hyperon, a different
linear combination should be used to ensure the stability of the sum
rules and subsequent results consistent with the experimental observation
\cite{Millener:1988hp, Noumi:2001tx}. The basis with scalar diquark
structure $u^T C\gamma_5 d$  has to be emphasized for this purpose.
Specifically, the ``stabilized'' interpolating fields for $\Lambda $
can be obtained by requiring $\tilde{a}\leq-2$ and $\tilde{b}\simeq
-0.2$ in the general expression~\eqref{lcrnt}. In this renewed
approach, the quasi-$\Lambda$ state is the light-bounded state
described by a weak scalar attraction and vector repulsion. The
strength of the self-energies is 30\% of the self-energies
calculated in the nucleon sum rules \cite{Cohen:1991js,
Furnstahl:1992pi, Jin:1992id, Jin:1993up, Cohen:1994wm,
Jeong:2012pa}. In the sum rules where the quasi-anti pole is given
as a constant calculated at the saturation
nuclear matter density, the
quasi-$\Lambda$ energy crosses with the quasineutron energy at
$\rho/\rho_0 \simeq 1.8$. In the case where the quasi-antipole is
calculated self consistently, the quasihyperon energy does not cross
with the quasineutron energy. As the linear density approximation
for the condensates is reliable only near the saturation
density, the sum-rule predictions are expected to be valid near the
saturation density region $0.5 <\rho/\rho_0<1.5$. To extend the
reliable results to the region $\rho/\rho_0>1.5$ the higher density
dependence in the condensate should be known. Thus, one can claim
that the onset of hyper-nuclear matter at low density would not
occur at least up to the density region $\rho/ \rho_0\leq 1.5$. The
large portion of the scalar diquark structure $u^T C\gamma_5 d$ in
the interpolating field  ensures the stability of the sum rules and
the acceptable density behavior of the quasi-$\Lambda$ state. Hence,
we expect that a good description of the  $\Lambda$ can be obtained
by introducing  a scalar field $\phi^\dagger_a$ to describe the two
light quarks as an single effective degree of freedom which contains
the quantum number of $\epsilon_{abc}[u_b^T C\gamma_5 d_c]$
($s^{P}=0^{+}$, $I=0$) as has been pursued in
Ref.~\cite{Kim:2011ut}.

The diquark structures of the interpolating fields for various baryons differ from each
other. For the nucleon case with Ioffe's choice \eqref{ncrnt1},
there is no unique way to make the diquark structure which consists
of two light quarks. If the two quarks are combined in different
flavor, scalar $u^T C\gamma_5 d$ and pseudoscalar $u^T C d$
structures are possible, but if the diquark is in the same flavor, it
should have nonzero spin structure ($q^T C\gamma_\mu q $ or $q^T
C\sigma_{\mu\nu} q $). Meanwhile, in the hyperon case, the diquark
structure can be restricted by the choice of the interpolating
fields and the sum-rules analysis. The diquark structure of Ioffe's
choice for $\Sigma$ hyperon \eqref{scrnt1} is the pseudovector $q_1^T
C\gamma_\mu q_2 $ where the light quark flavor $q_1$ and $q_2$ are
in the $I=1$ configuration. The parity condition requires the light quarks in these diquarks to be in $l=1$ state in the
nonrelativistic limit.  As for  the diquark structure $u^T C\gamma_5 d $ in the
stabilized $\Lambda$ interpolating field,  the quarks will be in the relative angular momentum $l=0$ state in the nonrelativistic limit.

Our sum rules show that the quasi-$\Lambda$ state is slightly
attractive and the quasi-$\Sigma$ state slightly repulsive in
neutron matter.   There seems to be a book keeping way of
understanding the interaction of the quasihyperon with the medium
in terms of the dominant diquark composed of light quarks.  Namely,
diquarks composed of lightquarks in the $l=0$ state are attractive
while those in the $l=1$ state are repulsive. It would be interesting
to investigate the validity of such a picture in models with
explicit diquark fields~\cite{Park:2016xrw}, whose property change
in nuclear medium can also be estimated in a constituent quark
picture~\cite{Park:2016xrw, Park:2016cmg}.  Such topics will be pursed
in a future work.

\begin{acknowledgements}
This work was supported by Korea National Research Foundation under the grant number KRF-2011-0030621 and the Korean ministry of education under the grant number 2016R1D1A1B03930089.  The
main
part of this work has been carried out  while KSJ was a BK-predoc fellow at
T30f/T39 group of Technical University of Munich. We thank Professor Nora Brambilla, Professor Norbert Kaiser, Professor Antonio Vairo and
Professor Wolfram Weise for the fruitful discussions and hospitality during the predoc program.
\end{acknowledgements}

\appendix

\section{Four-quark condensates}\label{appentw}

\subsection{Twist-4 matrix elements from deep inelastic scattering experiment}

\begin{table}
\addtolength{\tabcolsep}{+5pt}
\begin{tabular}{  c c c c  }
\hline\hline  Operator type & $\gamma-\gamma$ &
$\gamma_5\gamma-\gamma_5\gamma $ & $\sigma-\sigma$
\\
\hline $t^A-t^A$ & $\langle\bar{q}_1 \gamma_5 \gamma t^A q_1
\bar{q}_2 \gamma_5 \gamma t^A q_2\rangle_{p,\textrm{s.t.}}\equiv
T^1_{q_1q_2} $& $\langle\bar{q}_1  \gamma t^A q_1 \bar{q}_2  \gamma
t^A q_2\rangle_{p,\textrm{s.t.}}\equiv T^2_{q_1q_2} $ &
$\langle\bar{q}_1 \sigma t^A q_1 \bar{q}_2 \sigma
t^A q_2\rangle_{p,\textrm{s.t.}}\equiv T^5_{q_1q_2} $ \\[0.1cm]
$I-I$ & $\langle\bar{q}_1 \gamma_5 \gamma  q_1 \bar{q}_2 \gamma_5
\gamma   q_2\rangle_{p,\textrm{s.t.}}\equiv T^3_{q_1q_2} $&
$\langle\bar{q}_1  \gamma   q_1 \bar{q}_2  \gamma
q_2\rangle_{p,\textrm{s.t.}}\equiv T^4_{q_1q_2} $ &
$\langle\bar{q}_1 \sigma  q_1 \bar{q}_2 \sigma
  q_2\rangle_{p,\textrm{s.t.}}\equiv T^6_{q_1q_2} $ \\
\hline \hline
\end{tabular}\caption{Classification of the twist-4 operators and the corresponding matrix elements $T^i_{q_1q_2}$.}\label{t1}
\end{table}

In this section, the discussion for the twist-4 matrix elements in
Refs.~\cite{Choi:1993cu, Jeong:2012pa} will be extended. Also, the
consequence of this extension to sum-rule analysis will be
presented. We start from the proton expectation value of the following generic twist-4 operators \cite{Choi:1993cu, Jeong:2012pa}:
\begin{align}
\langle p \vert \bar{q}_{1} \Gamma^\alpha_i q_{1} \bar{q}_{2}
\Gamma^\beta_i q_{2}\vert p
\rangle_{\textrm{s.t.}}=\left(u^{\alpha}u^\beta
-\frac{1}{4}g^{\alpha\beta}\right)\langle  \bar{q}_{1} \Gamma _i
q_{1} \bar{q}_{2} \Gamma _i q_{2}  \rangle_{p,\textrm{s.t.}}
=\left(u^{\alpha}u^\beta
-\frac{1}{4}g^{\alpha\beta}\right)\frac{1}{4\pi\alpha_s}\frac{M_n}{2}T^i_{q_1q_2},
\end{align}
where $q_1,q_2$ represents the quark flavor. In this study, the
twist-4 operators which have strange quark flavor $\langle \bar{q}
\Gamma _i q \bar{s}  \Gamma _i s \rangle_{p,\textrm{s.t.}}$ are
omitted because the assumed nuclear matter in $0.5< \rho/\rho_0 <1.5$
will not allow for a larger external strange quark content. The
operator type $ \Gamma _i$ and the corresponding matrix elements
$T^i_{q_1q_2}$ can be found in Table~\ref{t1}.

\subsubsection{Twist-4 matrix elements for a single quark flavor}

From the relation~(84) of Ref.~\cite{Jeong:2012pa} and the ``zero
identity'' presented in Refs.~\cite{Thomas:2007gx, Jeong:2012pa},
the single quark flavored twist-4 operators whose matrix elements
had been estimated from deep inelastic scattering (DIS) experiment \cite{Choi:1993cu,
Jeong:2012pa} can be summarized as
\begin{align}
[\bar{q}\gamma^{\alpha} t^A q\bar{q}\gamma^{\beta} t^A q]_{\textrm{s.t.}}=&-\frac{5}{12}[\bar{q}\gamma^{\alpha} q \bar{q} \gamma^{\beta} q]_{\textrm{s.t.}} -\frac{1}{4}
[\bar{q}\gamma_5 \gamma^{\alpha} q \bar{q} \gamma_5\gamma^{\beta} q] _{\textrm{s.t.}} +\frac{1}{4} [\bar{q} \sigma_{\mu}^{~\alpha} q \bar{q}
\sigma^{\mu\beta} q ]_{\textrm{s.t.}}, \label{cond1}\\
[\bar{q}\gamma_5\gamma^{\alpha} t^A q \bar{q} \gamma_5\gamma^{\beta} t^A q ]_{\textrm{s.t.}}=&-\frac{5}{12}[\bar{q}\gamma_5 \gamma^{\alpha} q \bar{q}\gamma_5 \gamma^{\beta}
q]_{\textrm{s.t.}} -\frac{1}{4}[\bar{q}\gamma^{\alpha} q \bar{q} \gamma^{\beta} q ]_{\textrm{s.t.}} -\frac{1}{4} [\bar{q} \sigma_{\mu}^{~\alpha} u \bar{q} \sigma^{\mu\beta} q
]_{\textrm{s.t.}}. \label{cond2}
\end{align}
The matrix elements for the operators on the left-hand sides will be
denoted $T^1_{qq}$ and $T^2_{qq}$ as given in
Table~\ref{t1}. By using successive the Fierz rearrangement, the
operator in $T^5_{qq}$ can be written as follows:
\begin{align}
[\bar{q} \sigma_{\mu}^{~\alpha} t^A q \bar{q} \sigma^{\mu\beta} t^A q ]_{\textrm{s,t}}=&~ \frac{1}{2}[\bar{q}\gamma^{\alpha} q \bar{q} \gamma^{\beta} q]_{\textrm{s,t}}
-\frac{1}{2}[\bar{q}\gamma_5 \gamma^{\alpha} q \bar{q} \gamma_5\gamma^{\beta} q] _{\textrm{s,t}} -\frac{1}{6} [\bar{q} \sigma_{\mu}^{~\alpha} q \bar{q} \sigma^{\mu\beta} q
]_{\textrm{s,t}}, \label{cond3}
\end{align}
which leads to the following relation:
\begin{align}
\epsilon_{abc}\epsilon_{a'b'c}[\bar{q}_{a'} \sigma q_{a} \bar{q}_{b'} \sigma q_{b} ]_{\textrm{s,t}}=&~\frac{2}{3}[\bar{q} \sigma_{\mu}^{~\alpha} q \bar{q} \sigma^{\mu\beta}
q ]_{\textrm{s.t.}}-2[\bar{q} \sigma_{\mu}^{~\alpha} t^A q \bar{q} \sigma^{\mu\beta} t^A q
]_{\textrm{s.t.}}\nonumber\\
=&- [\bar{q}\gamma^{\alpha} q \bar{q} \gamma^{\beta} q]_{\textrm{s.t.}} + [\bar{q}\gamma_5 \gamma^{\alpha} q \bar{q} \gamma_5\gamma^{\beta} q] _{\textrm{s.t.}} + [\bar{q}
\sigma_{\mu}^{~\alpha} q \bar{q} \sigma^{\mu\beta} q ]_{\textrm{s.t.}}.
\end{align}
The corresponding matrix elements can be expressed as
\begin{align}
 \langle \bar{q} \sigma q
\bar{q} \sigma q \rangle _{\textrm{s,t}} =&~\frac{1}{4\pi\alpha_s}\frac{M_n}{2}\bigg[- \frac{1}{2}([T^4_{uu}+T^4_{dd}] \mp [T^4_{uu}-T^4_{dd}]I ) +
\frac{1}{2}([T^3_{uu}+T^3_{dd}] \mp [T^3_{uu}-T^3_{dd}]I ) \nonumber\\
&\qquad\qquad\quad + \frac{1}{2}([T^6_{uu}+T^6_{dd}] \mp [T^6_{uu}-T^6_{dd}]I )\bigg]\rho, \label{cond4}
\end{align}
where  ``$+$'' and ``$-$'' stand for $u$ and $d$ quark flavor,
respectively. In our previous work \cite{Jeong:2012pa}, the matrix
elements for the operator
$\epsilon_{abc}\epsilon_{a'b'c}[\bar{q}_{a'} \sigma q_{a}
\bar{q}_{b'} \sigma q_{b} ]_{\textrm{s,t}}$ were neglected because the
contribution of these operators to the DIS process was expected to be minimal
in Ref.~\cite{Jaffe:1981sz}. However, as one can find in
Eq.~\eqref{cond4}, in the case where $T^3_{qq} \neq 0$ and $T^4_{qq}
\neq 0$, the matrix elements of $\langle \bar{q}  \sigma  q \bar{q}
\sigma q  \rangle _{\textrm{s,t}}$ cannot automatically be zero. In
this study, we take the matrix elements $T^6_{qq} $ as free
parameters and estimate their value. One needs
additional assumption for the ratio  $T^6_{uu}/T^6_{dd} $.
As for $T^1_{qq}$ and $T^2_{qq}$, the ratios between $u$ and $d$ flavors  can be  assumed to be $T^1_{uu}/T^1_{dd} = T^2_{uu}/T^2_{dd}\simeq 6$
\cite{Jeong:2012pa} from estimates of DIS experiment.   Based on this observation, we will also
take the ratio of
 $T^6_{uu}/T^6_{dd} \Rightarrow 6$. Then, $T^3_{qq}$ and $T^4_{qq}$ can be summarized as
follows:
\begin{align}
T^3_{qq}=-\frac{15}{4}T^1_{qq}+\frac{9}{4}T^2_{qq}-\frac{3}{2}T^6_{qq},\\
T^4_{qq}=-\frac{15}{4}T^2_{qq}+\frac{9}{4}T^1_{qq}+\frac{3}{2}T^6_{qq}.
\end{align}
The single quark flavored matrix elements are listed in
Table~\ref{t2}.

\begin{table}
\addtolength{\tabcolsep}{+8pt}
\begin{tabular}{  c c c c c c c c }
\hline\hline $T^1_{uu}$ & $T^1_{dd}$ & $T^2_{uu}$& $T^2_{dd}$ &
$T^3_{uu}$ & $T^3_{dd}$ & $T^4_{uu}$ & $T^4_{dd}$
\\
\hline $-0.071$ & $-0.012$ & $0.070$& $0.012$ &
$0.424-\frac{3}{2}T^6_{uu}$ &$0.072-\frac{3}{2}T^6_{dd}$
&$-0.424+\frac{3}{2}T^6_{uu}$ &$-0.072+\frac{3}{2}T^6_{dd}$ \\
\hline \hline
\end{tabular}\caption{Table for the twist-4 matrix elements for a single quark flavor $T^i_{qq}$. Units are $\textrm{GeV}^2$.}\label{t2}
\end{table}

\subsubsection{Twist-4 matrix elements for  mixed quark
flavor}

The matrix element $T^1_{ud}=-0.042~\textrm{GeV}^2$ had been
uniquely determined by DIS experiment \cite{Choi:1993cu}. Using the
arguments in Refs.~\cite{Choi:1993cu, Jeong:2012pa},
$T^2_{ud}=0.041~\textrm{GeV}^2$ has also been estimated. For
$T^1_{uu}$ and $T^1_{dd}$, one observes the relation
$T^1_{ud}\simeq (T^1_{uu}+T^1_{dd})/2=-0.042~\textrm{GeV}^2$.
Similarly, $T^2_{ud}\simeq
(T^2_{uu}+T^2_{dd})/2=0.041~\textrm{GeV}^2$. Based on this
observation, we claim that the other matrix elements also satisfy
the same relation: $T^3_{ud}\simeq (T^3_{uu}+T^3_{dd})/2$ and
$T^4_{ud}\simeq (T^4_{uu}+T^4_{dd})/2$. Then, for the mixed quark
flavored case, the relation \eqref{cond4} can be rewritten as
follows:
\begin{align}
 \langle \bar{u} \sigma  u
\bar{d}\sigma d\rangle _{\textrm{s.t.}}
=~\frac{1}{4\pi\alpha_s}\frac{M_n}{2}\left(- T^4_{ud}
   + T^3_{ud}  + \frac{7}{12}T^6_{uu}   \right)\rho,\label{cond5}
\end{align}
where the iso-spin-dependent pieces have been excluded by assumed
symmetries. The mixed quark flavored matrix elements are listed in
Table~\ref{t3}.

\begin{table}
\addtolength{\tabcolsep}{+32pt}
\begin{tabular}{  c c c c  }
\hline\hline $T^1_{ud}$ & $T^2_{ud}$ & $T^3_{ud}$ & $T^4_{ud}$
\\  \hline $-0.042$ &   $0.041 $& $T^3_{ud}\simeq
0.250-\frac{3}{2}T^6_{ud}$ & $T^4_{ud}\simeq
-0.250+\frac{3}{2}T^6_{ud}$
 \\
\hline \hline
\end{tabular}\caption{Table for the twist-4 matrix elements for mixed quark flavor $T^i_{ud}$. $T^6_{ud}=(7/12)T^6_{uu}$. Units are $\textrm{GeV}^2$.}\label{t3}
\end{table}

\subsection{Parameter dependence of the sum rule analysis}

\begin{figure}
\includegraphics[height=6.5cm]{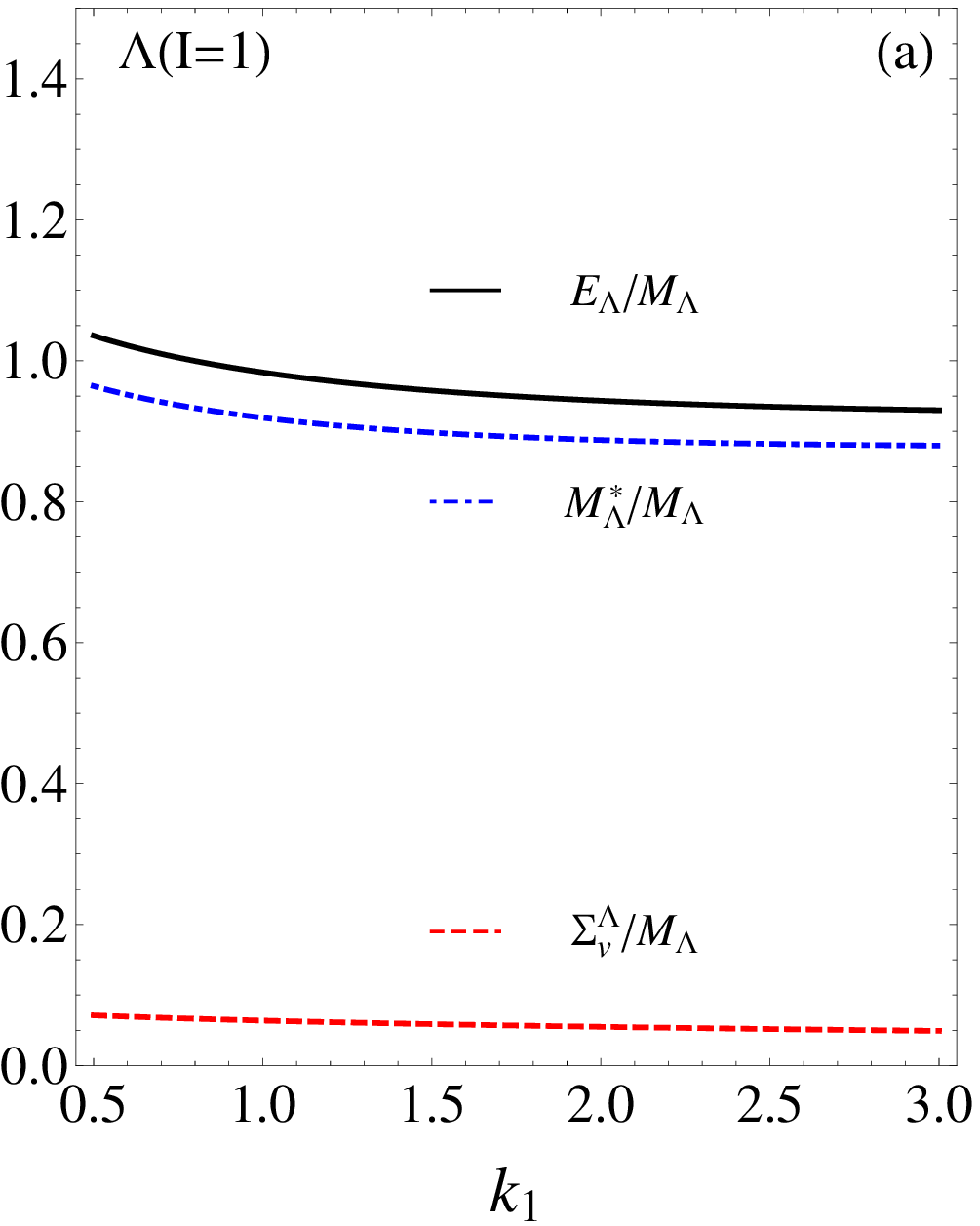}
\includegraphics[height=6.5cm]{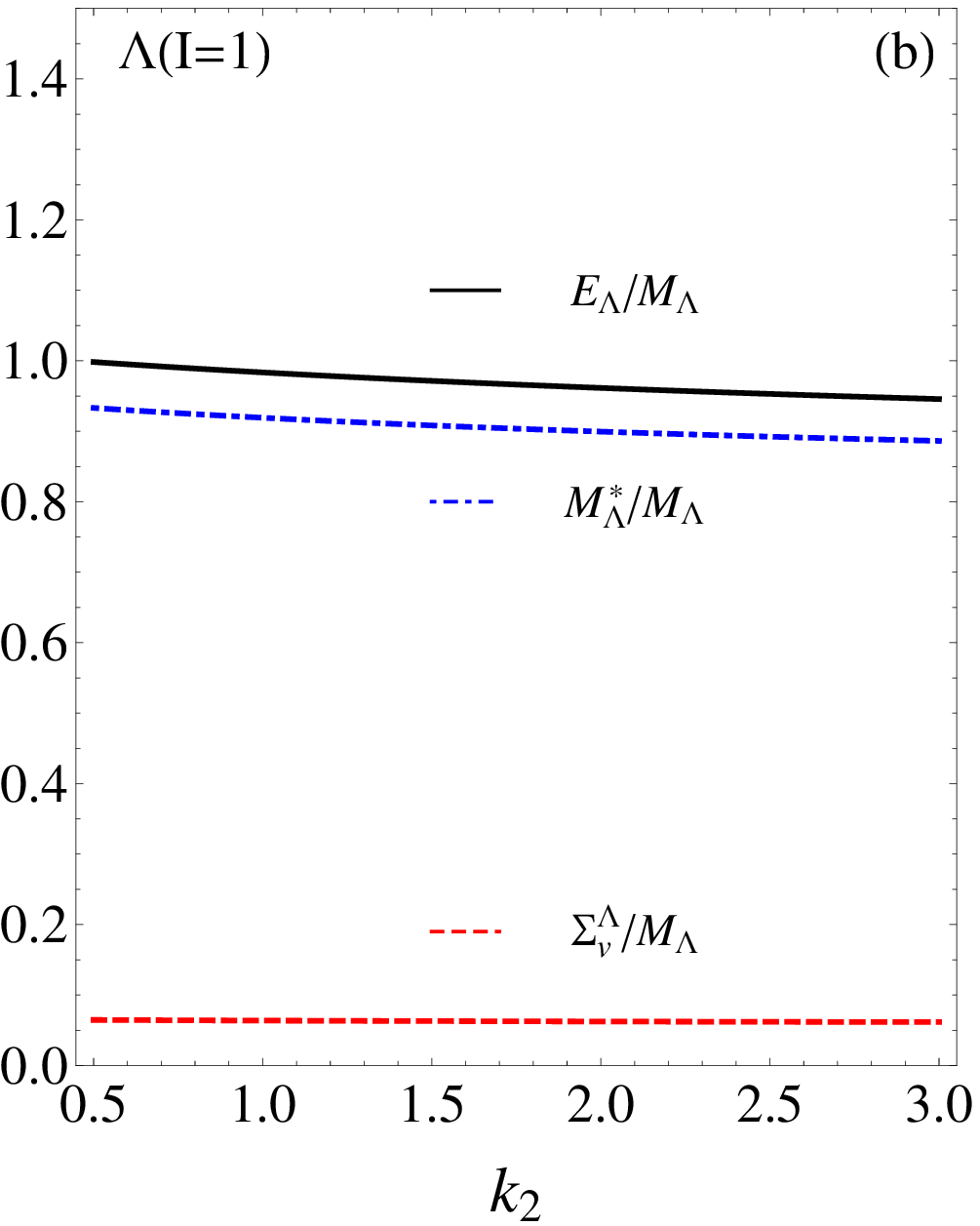}
\includegraphics[height=6.5cm]{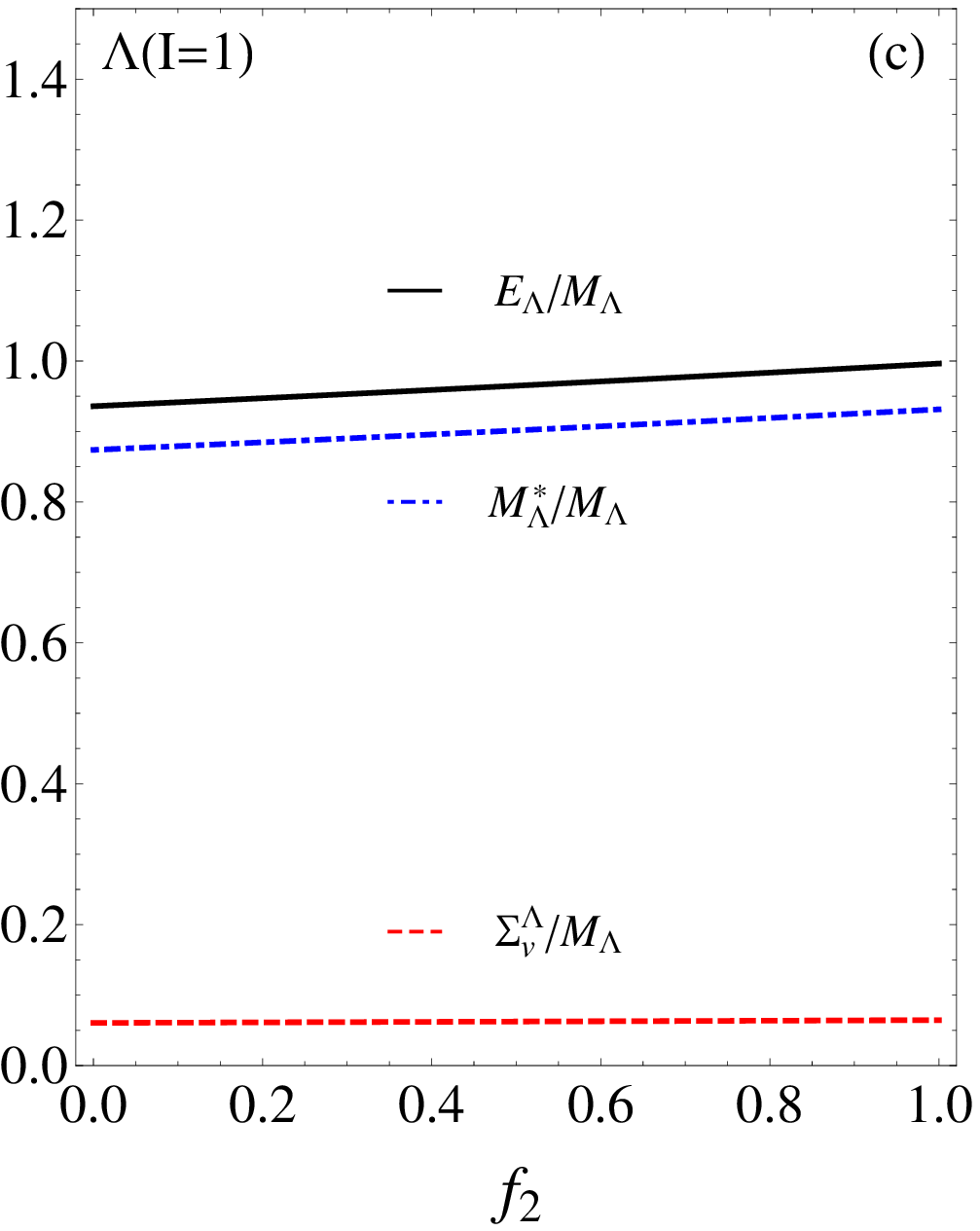}
\caption{Factorization parameter dependence of the
sum rule for the quasi-$\Lambda$ state ($\rho=\rho_0$,
$M^2=1.1~\textrm{GeV}^2$). (a) $k_1$ dependence from the
factorization $\langle \bar{q}q \bar{q}q \rangle_{\textrm{vac}}
\Rightarrow k_1\langle\bar{q}q\rangle^2_{\textrm{vac}}$. (b) $k_2$
dependence
 from the
factorization $\langle \bar{q}q \bar{s}s \rangle_{\textrm{vac}}
\Rightarrow
k_2\langle\bar{q}q\rangle_{\textrm{vac}}\langle\bar{s}s\rangle_{\textrm{vac}}$.
(c) $f_2$ dependence from the factorization $\langle \bar{q}q
\bar{s}s \rangle_{\textrm{med.}} \Rightarrow f_2\left(
\langle\bar{s}s \rangle_p
\langle\bar{q}q\rangle_{\textrm{vac}}+\langle[\bar{q}q]_0 \rangle_p
\langle\bar{s}s\rangle_{\textrm{vac}}\right)\rho$.}\label{pr}
\end{figure}

First, we examine the factorization parameter
dependence of the sum rules. In Fig.~\ref{pr}, the $k_1$, $k_2$, and
$f_2$ dependence of the quasi-$\Lambda$ self-energies are plotted.
As can be seen in  Fig.~\ref{pr}, the dependencies on the parameters $k_1$, $k_2$, and
$f_2$ are weak in the quasi-$\Lambda$ sum rules.
Hence, we
have chosen the parameters as used in the previously reported
studies \cite{Shifman:1978bx, Ioffe:1981kw, Reinders:1984sr,
Drukarev:1988kd, Cohen:1991js, Hatsuda:1991ez, Furnstahl:1992pi,
Jin:1992id, Jin:1993up, Cohen:1994wm, Jeong:2012pa}:
$k_1,k_2,f_2=1$. While we chose $k_1,k_2=1$, the sum-rule results do not change much even if the
scalar four-quark condensates are taken to be $50\% \sim 300\%$ of the
estimated value.

\begin{figure}
\includegraphics[height=6.5cm]{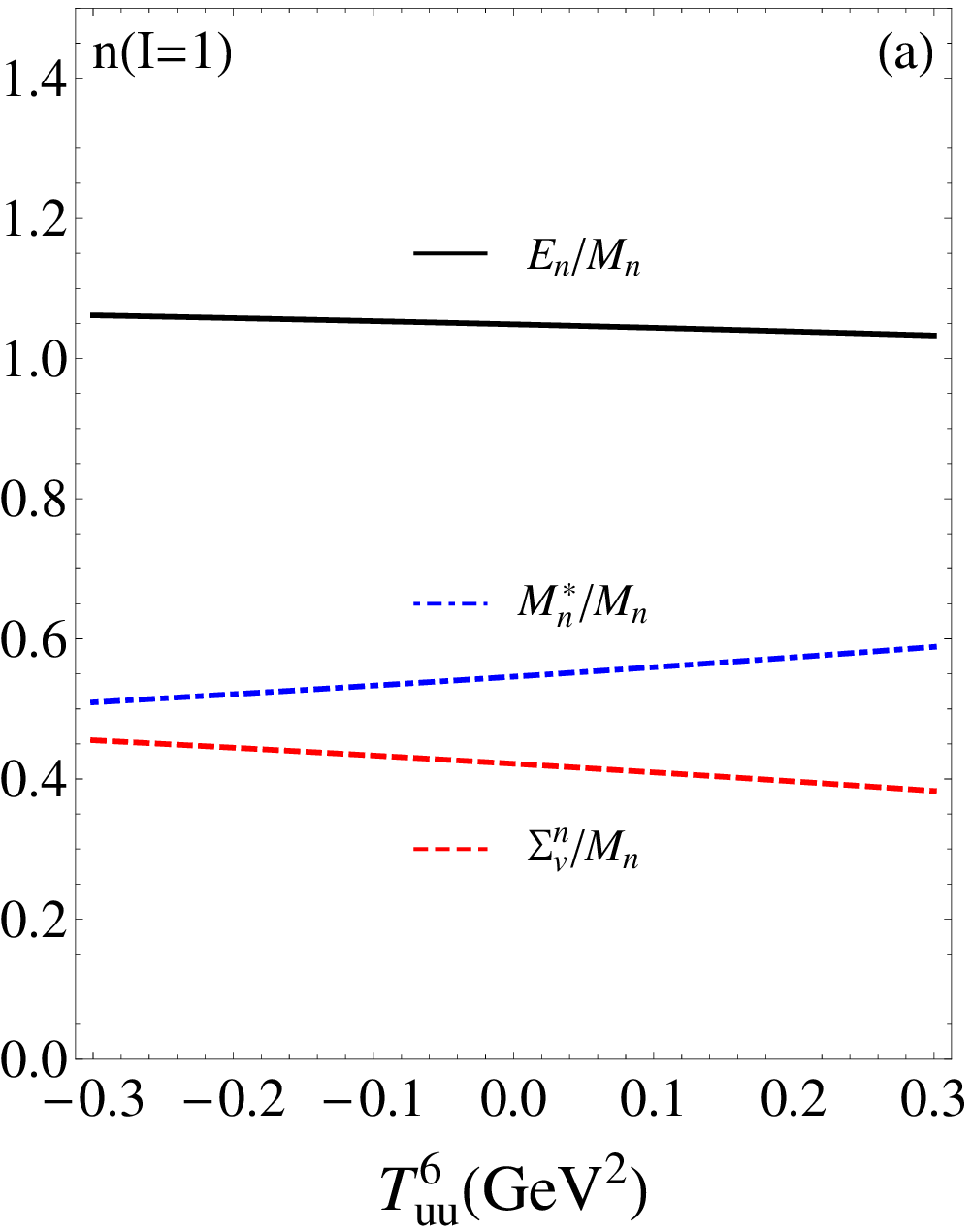}
\includegraphics[height=6.5cm]{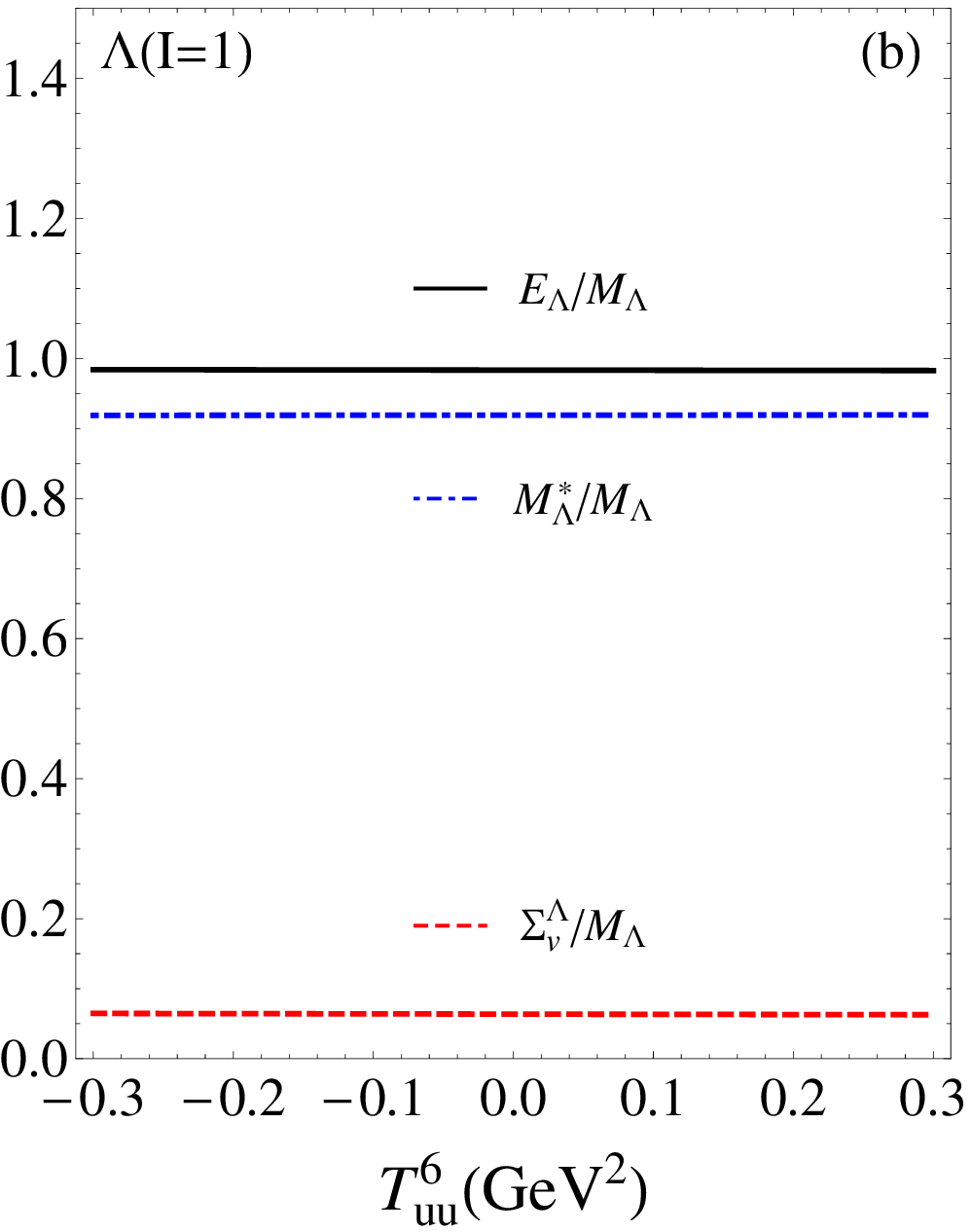}
\includegraphics[height=6.6cm]{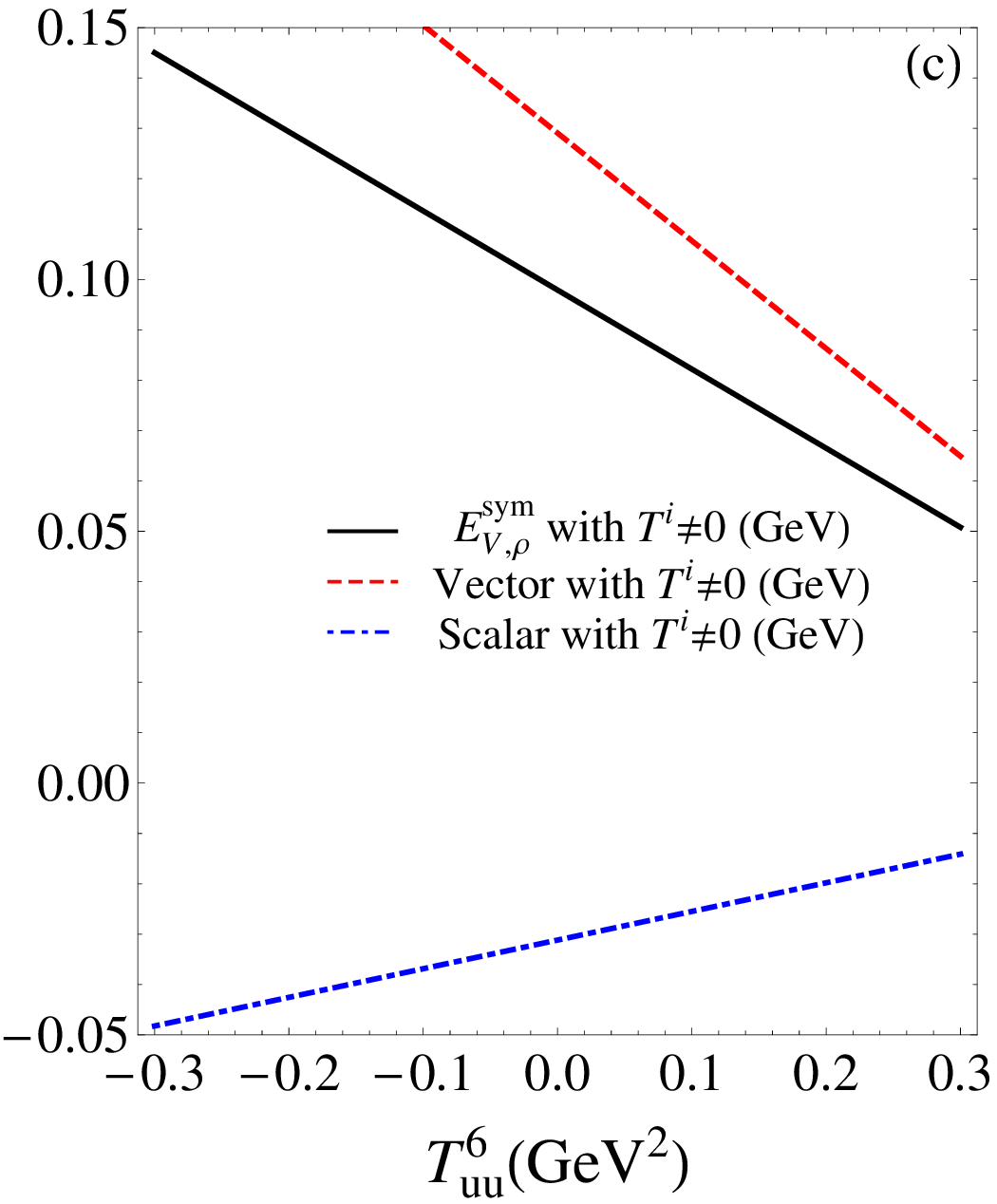}
\caption{Sum rules with various $T^6_{uu}$
($\rho=\rho_0$). (a) the quasineutron sum rules in the neutron
matter. (b) the quasi-$\Lambda$ sum rules in the neutron matter. (c)
nuclear symmetry energy. The OPE terms for the symmetry energy can
be found in Ref.~\cite{Jeong:2012pa}. Units for graph (c) are
GeV.}\label{t6}
\end{figure}

The influences of $T^6_{uu}$ are plotted in Fig.~\ref{t6}. As can be
found in Fig.~\ref{t6}(a), the quasi-neutron energy has weak
dependence on $T^6_{uu}$ but each self-energy has non-negligible
dependence. The scalar (vector) self-energy becomes enhanced
(reduced) as $T^6_{uu}$ grows whereas the quasi-$\Lambda$ sum rules
are almost independent on $T^6_{uu}$ [Fig.~\ref{t6}(b)]. The
stabilized sum rules with the interpolating fields require very
small $\tilde{b}$, which multiplies the twist-4 condensates term in
the OPE. Hence, the contribution of the twist-4 condensates is
minimal regardless of the value of matrix elements parametrized by
$T^6_{uu}$ because $\tilde{b}$ is small. In Fig.~\ref{t6}(c), the nuclear
symmetry energy is parametrized by $T^6_{uu}$. The symmetry energy
becomes reduced and the cancellation mechanism between the vector and
scalar becomes weaker as $T^6_{uu}$ grows. We have chosen
$T^6_{uu}=0.2~\textrm{GeV}^2$ to ensure moderate strength of the
cancelation mechanism shown in Figs.~\ref{t6}(a) and~\ref{t6}(c).

\end{document}